\documentclass[12pt]{article}
\usepackage{geometry}
\usepackage{amsmath}
\usepackage{amsfonts}          
\usepackage{amssymb} 
\usepackage{xspace}  
\usepackage{graphicx}   
\usepackage{color} 
\usepackage{xy}
\usepackage{rotating}

\def\appendix#1{
  \addtocounter{section}{1}
  \setcounter{equation}{0}
  \renewcommand{\thesection}{\Alph{section}}
 \section*{Appendix \thesection\protect\indent \parbox[t]{11.715cm} {#1}}
  \addcontentsline{toc}{section}{Appendix \thesection\ \ \ #1}
  }

\renewcommand{\thefootnote}{\fnsymbol{footnote}}

\numberwithin{equation}{section}

\newcommand{\be}{\begin{equation}}
\newcommand{\ee}{\end{equation}}
\newcommand{\ba}{\begin{aligned}}
\newcommand{\ea}{\end{aligned}}

\newcommand{\ie}{{\it i.e.}}

\newcommand{\p}{\partial}

\newcommand{\hatw}{{\hat{w}}}
\newcommand{\hatv}{{\hat{v}}}
\newcommand{\hatx}{{\hat{x}}}
\newcommand{\vect}{\mathrm{Vect}}
\newcommand{\hatSigma}{\hat{\Sigma}}
\newcommand{\hatM}{\hat{M}}
\newcommand{\ft}{\mathrm{F.T.}}

\makeatletter
\def\sla@#1#2#3#4#5{{%
  \setbox\z@\hbox{$\m@th#4#5$}%
  \setbox\tw@\hbox{$\m@th#4#1$}%
  \dimen4\wd\ifdim\wd\z@<\wd\tw@\tw@\else\z@\fi
  \dimen@\ht\tw@
  \advance\dimen@-\dp\tw@
  \advance\dimen@-\ht\z@
  \advance\dimen@\dp\z@
  \divide\dimen@\tw@
  \advance\dimen@-#3\ht\tw@
  \advance\dimen@-#3\dp\tw@
  \dimen@ii#2\wd\z@
  \raise-\dimen@\hbox to\dimen4{%
    \hss\kern\dimen@ii\box\tw@\kern-\dimen@ii\hss}%
  \llap{\hbox to\dimen4{\hss\box\z@\hss}}}}
\def\slashed#1{%
  \expandafter\ifx\csname sla@\string#1\endcsname\relax
    {\mathpalette{\sla@/00}{#1}}%
  \else
    \csname sla@\string#1\endcsname
  \fi}
\makeatother

\begin{document}


\thispagestyle{empty}
\begin{flushright}\footnotesize
\texttt{arxiv:0708.2392}\\
\texttt{CALT-68-2658}\\
\texttt{DESY 07-127}\\
\texttt{ZMP-HH/07-022}\\
\vspace{2.1cm}
\end{flushright}

\renewcommand{\thefootnote}{\fnsymbol{footnote}}
\setcounter{footnote}{0}
\begin{center}
{\Large\textbf{\mathversion{bold} 
Towards mirror symmetry \emph{\`a  la} SYZ\\
for generalized Calabi--Yau manifolds
}\par}
\vspace{2.1cm}

\textrm{Pascal Grange$\,^\ast$ and Sakura Sch\"afer-Nameki$\,^\sharp$}

\vspace{1cm}

\textit{$^\ast$ II. Institut f\"ur theoretische Physik der Universit\"at Hamburg\\
Luruper Chaussee 149, 22761 Hamburg, Germany} \\

\vspace{1mm}
\textit{and}
\vspace{1mm}

\textit{Zentrum f\"ur mathematische Physik, Universit\"at Hamburg\\
Bundesstrasse 55, 20146 Hamburg, Germany} 

\vspace{5mm}

\textit{$^\sharp$ California Institute of Technology\\
1200 E California Blvd., Pasadena, CA 91125, USA
}

\vspace{6mm}
\texttt{pascal.grange@desy.de, ss299@theory.caltech.edu}

\vspace{3mm}


\par\vspace{1cm}

\textbf{Abstract}
\end{center}

\noindent
Fibrations of flux backgrounds by supersymmetric cycles are investigated. For an internal six-manifold $M$ with static $SU(2)$ structure and mirror $\hat{M}$, it is argued that the product $M\times \hat{M}$ is doubly  fibered by supersymmetric three-tori, with both sets of fibers transverse to $M$ and $\hat{M}$. The mirror map is then realized by T-dualizing the fibers. Mirror-symmetric properties of the fluxes, both geometric and non-geometric, are shown to agree with previous conjectures based on the requirement of mirror symmetry for Killing prepotentials. The fibers are conjectured to be destabilized by fluxes on generic  $SU(3)\times SU(3)$ backgrounds, though they may survive at type-jumping points. T-dualizing the surviving fibers ensures the exchange of pure spinors under mirror symmetry.

\vspace{5mm}

\vspace*{\fill}

\newpage
\setcounter{page}{1}
\renewcommand{\thefootnote}{\arabic{footnote}}
\setcounter{footnote}{0}


\tableofcontents


\section{Introduction}

The study of flux compactifications is strongly motivated by the
necessity to fix the moduli of the compact space. It leads to
the consideration of flux backgrounds which lack certain geometric features of
Calabi--Yau manifolds: typically the closure of the two- and three-forms
of Calabi--Yau manifolds are spoiled by intrinsic torsion. 
Moreover, the duality symmetries of string theory lead to backgrounds that are
non-geometric in the sense that the closed-string metric is not
globally defined. This concept appeared first in various incarnations in \cite{Kachru:2002sk, Hellerman:2002ax, Mathai:2004qq, Hull:2004in, Shelton:2005cf} 
and a unifying picture connecting these various points of view was proposed using generalized geometry in \cite{Grange:2006es}.

 However, there is still some structure surviving in flux backgrounds
 preserving eight supercharges in four dimensions: such backgrounds
 have to possess $SU(3)\times SU(3)$ structure \cite{Jeschek:2004wy, Grana:2005ny,
 Grana:2006hr}. This implies the existence of a pair of pure spinors
 of different parity $\Phi_+$ and $\Phi_-$, one being closed and
 inducing a generalized complex structure, so that the internal space
 is a generalized Calabi--Yau manifold \cite{Hitchin:2004ut,
 Gualtieri:2003dx}.  The other one is not closed in the presence of
 Ramond-Ramond fluxes, but its imaginary part is, and gives rise to
 calibrations \cite{Gukov:1999ya,Grana:2005sn,Koerber:2005qi,Martucci:2005ht,Gmeiner:2006ni,Koerber:2006hh}.

 $SU(3)$-structures form a geometric subclass of $SU(3)\times SU(3)$ structure manifolds, where
the pure spinors are denoted by $\Omega$ and $e^{i J}$, and there is no type-jumping. These manifolds were established to be the mirrors of Calabi-Yau with so-called geometric or electric $H$--flux in \cite{Gurrieri:2002wz}, which in the case of torus-bundles reduces to the statement that T-duality exchanges the Chern-class of the bundle with the integral of the $H$-flux along the T-dualized direction \cite{Kachru:2002sk, Alvarez:1993qi, Cavalcanti:2005hq}.

 Special cases of non-geometric backgrounds have been identified as
 physical realizations of the type-jumping phenomenon previously
 studied in generalized complex geometry \cite{Gualtieri:2003dx,Grange:2006es,Micu:2007rd}. 
 Furthermore, the effective
 actions of string theory on backgrounds admitting $SU(3)\times SU(3)$ 
 structure exhibit symmetry properties under the exchange of $\Phi_+$
 and $\Phi_-$ \cite{Grana:2006hr}. This exchange extends the action of mirror symmetry
 beyond the realm of Calabi--Yau manifolds, in which the pure spinors are $\Phi_+= e^{iJ}$
 and $\Phi_-= \Omega$, where $J$ and $\Omega$ denote the K\"ahler form
 and holomorphic three-form, respectively. The generalized
 calibrations are exchanged in the same way as the ones governing
 stability of D-branes of type A and B on Calabi--Yau manifolds \cite{Grange:2004ah,Martucci:2005ht}.
 Fortunately generalized complex submanifolds share a lot of properties with Abelian D-branes \cite{BB1,Grange:2005nm,Evslin:2007ti}.

 Flux backgrounds, while fixing moduli, have therefore violently
 shaken the geometric framework of Calabi--Yau compactifications, but
 still happen to possess good mirror-symmetric properties. This begs
 for an explanation in terms of the action of T-duality on the
 internal space in the presence of fluxes. In other words, we would
 like to know what remains of the Strominger--Yau--Zaslow (SYZ) picture  of
 mirror symmetry \cite{Strominger:1996it, MR1876066}, in the case of $SU(3)\times
 SU(3)$ structure backgrounds.

  The purpose of this paper is therefore to investigate the moduli
  space of calibrated cycles in backgrounds with $SU(3)\times SU(3)$
  structure, and to formulate the exchange between $\Phi_+$ and
  $\Phi_-$ in terms of T-duality along such cycles, thus extending
  mirror symmetry to cases where much of the structure available in
  Calabi--Yau manifolds is missing.\footnote{For generalized K\"ahler manifolds an argument of mirror symmetry via T-duality for the topological sigma-models was put forward in \cite{Chiantese:2004tx}.}
 
 This can be done in several steps. After recalling the connection
 between the pure spinors and the supercharges, we specialize to
 the case of internal manifolds with a so-called static $SU(2)$
 structure. The type of the pure spinors are constant on such
 manifolds, but never maximal, since they are equal to one and two, respectively. We
 shall see that supersymmetric tori transverse to the product of the
 internal space $M$ and its mirror $\hat{M}$ have the entire $M\times
 \hat{M}$ as moduli space. In particular $M$ and $\hat{M}$ are still
 fibered by three-tori, but the fibers are not supersymmetric by
 themselves. We illustrate this generalized SYZ proposal for static $SU(2)$ structure manifolds in various special cases and show that it is compatible with the mirror map advocated in \cite{Grana:2006hr}.

 Then we address the case of generic $SU(3)\times SU(3)$
 structures, that exhibit type-jumping phenomena, and correspondingly
 open-string moduli fixing. We shall see that zeroes or critical
 points of the coefficients relating the supercharges to each other
 dictate the position moduli of supersymmetric cycles. Finally, we may
 perform T-duality along the existing supersymmetric cycles, and
 obtain the type-jumping phenomena from the naturality properties of
 Fourier--Mukai transform with respect to the so-called $B$- and
 $\beta$-transforms of generalized complex geometry. This will be
 related to the covariance properties of the differential operators on
 flux backgrounds, and confirm the mirror-symmetric form of the
 superpotentials for $SU(3)\times SU(3)$ backgrounds.


\section{Review and notations}

\subsection{Supersymmetry, pure spinors and structures}

Generalized complex geometry contains both complex geometry and symplectic geometry. An almost generalized complex structure on a manifold $M$ is defined as an almost complex structure on the sum of the tangent and cotangent bundles. It is a generalized complex (GC) structure  if its $+i$-eigenbundle is stable under the action of the Courant bracket \cite{Hitchin:2004ut,Gualtieri:2003dx,Ellwood:2006ya}. We will give a more detailed review of the concepts in generalized geometry, including GC submanifolds, in the next sub-section. 
Here we review the definition of pure spinors in terms of supercharges. 
There is a one-to-one correspondence between GC structures and pure spinors. A pure spinor is a sum of differential forms and may locally be written in a unique way as the wedge product of $k$ complex one-forms and the exponential of a two-form:
\be
\theta_1\wedge\dots\wedge\theta_k \wedge e^{B+i\omega}.
\ee  
The integer $k$ is called the \emph{type} of the pure spinor. From now on we only consider six-dimensional manifolds. The special case $k=0$ corresponds to a symplectic structure on the manifold, and the special case $k=3$ to a complex structure. Not only can the type assume other values, but it can also vary on the manifold. This is called the {\emph{type-jumping}} phenomenon \cite{Gualtieri:2003dx}. We will mostly work with the pure spinors as the objects encoding the GC structure.

Consider Type II compactifications on six-manifolds with $SU(3)\times SU(3)$ structure \cite{Hitchin:2004ut,Gualtieri:2003dx,Jeschek:2004wy,Grana:2005ny,Grana:2006hr} (for more references see \cite{Grana:2005jc}). 
These are characterized by a pair of no-where vanishing $SU(3)$-invariant spinors $\eta^{1,2}$, 
which arise in the decomposition of the two $SO(9,1)$ spinors $\epsilon^{1,2}$ of Type II under 
$SO(3,1) \times SO(6)$. 

Let $M$ and $\hat{M}$ be a (real) six-dimensional manifold and its
mirror, both assumed to have $SU(3)\times SU(3)$ structure. As such
they respectively possess pure spinors $\Phi_-,\Phi_+$ and
$\hat{\Phi}_-,\hat{\Phi}_+$, where the signs denote the parity of the
type. The pure spinors on $M$ are constructed as bilinears of spinors:
\be
\ba
\Phi_+  &=\eta_+^1\otimes \eta_+^{2\dagger} \cr
\Phi_-   &=\eta_+^1\otimes \eta_-^{2\dagger} \,,
\ea
\ee
where $\eta^1$ and $\eta^2$ are related to each
other by the equation 
\be
\eta_+^2=c\eta_+^1+(v+iw)_m\gamma^m\eta_-^1\,,
\ee
defining the complex one-form $v+iw$  and complex number $c$, which have to satifsy the
normalization condition 
\be
|c|^2+|v+iw|^2=1\,.
\ee 
There are analogous
objects on $\hat{M}$ and we shall occasionally refer to them just by putting hats on the symbols we explicitly define on $M$.

\begin{figure}[th]
\centerline{\includegraphics[width=11cm]{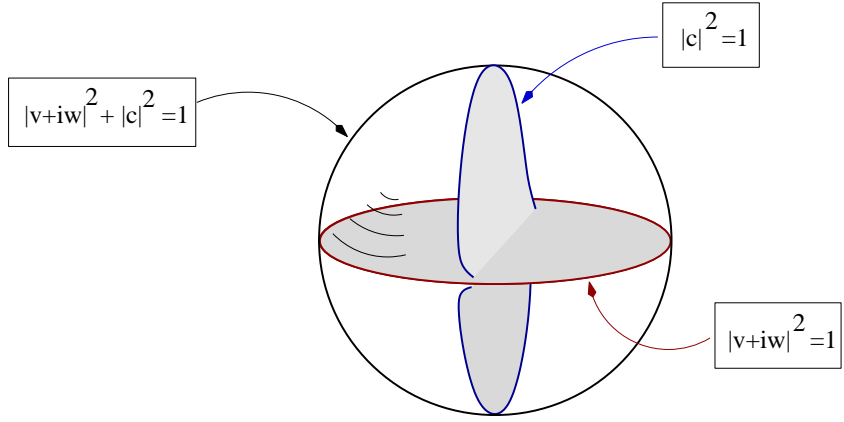}}
\caption{\label{fig1}  
Moduli space of spinors for $SU(3)\times SU(3)$ structure manifolds (depicted one dimension lower as an $S^2$). The blue and red circles depict the $SU(3)$ and the static $SU(2)$ structures, respectively, which do not intersect.  
}
\end{figure}

 At points where $|c|=1$ (zeroes of $v+iw$), the two $SU(3)$
 spinors $\eta^1$ and $\eta^2$ become proportional to each other, and
 the two $SU(3)$ structures defined by bilinears of $\eta^1$ and
 $\eta^2$ agree. At such points the pure spinors $\Phi_-$ and $\Phi_+$
 have type three and type zero respectively, just as they do in the
 case of manifolds of $SU(3)$ structures. 
Up to a $B$-transform they
 read at such points 
 \be\label{SU(3)Structure}
 \ba
 {\Phi_-}|_{|c|=1} &=\Omega\cr
 {\Phi_+}|_{|c|=1} &= e^{iJ} \,.
 \ea
 \ee

 If $|c|=1$ on the whole of $M$, then $M$ has an $SU(3)$ structure. 
 Calabi--Yau manifolds form the subclass of those manifolds for which both of $\Omega$ and $e^{iJ}$ are closed. 

 At generic points though, the spinors
 $\eta^1$ and $\eta^2$ are linearly independent, the two $SU(3)$
 structures constructed from them do not agree, and their fundamental
 two-form and complex three-form may be written as 
 \be \label{JJDef}
 \ba
 J^1 =j+v\wedge w \,,\qquad  &J^2=j-v\wedge w \cr
 \Omega^1 =\omega\wedge (v+iw)\,,\qquad 
 &\Omega^2=\omega\wedge (v-iw) \,.
 \ea
 \ee
The pure spinors in turn are expressed \cite{Jeschek:2004wy} in a way that allows to read-off their types, as
\be
\ba
\Phi_-  & =-\frac{1}{8}\,(v+iw)\wedge e^{i(j+c\omega)} \cr
\Phi_+ &= -\frac{1}{8}\,\bar{c}\, e^{-i(j+v\wedge w +\frac{1}{\bar{c}}\omega)}\,.
\ea
\ee
It can be observed that type-jumping (from one to three) occurs for $\Phi_-$ at points where $|c|=1$ (zeroes of $v+iw$). 
 The limit where $v+iw$ goes to zero is ill-defined in those expressions, and the pure spinors at such points are expressed as in formulas (\ref{SU(3)Structure}).

 Another type-jumping phenomenon occurs at zeroes of $c$. At these
 points the two spinors $\eta^1$ and $\eta^2$ become orthogonal, and
 there is a local $SU(2)$ structure. The pure spinors then read:
\be
\ba 
 \Phi_- &= (v+iw)\wedge e^{ij} \cr
\Phi_+ &= \omega\wedge e^{i{v\wedge w}} \,.
\ea
\ee

We notice that due to the normalization constraint relating $c$ to
$v+iw$, type-jumping occurs at critical points of $|c|$ and
$|v+iw|$. 
The situation is depicted in Figure \ref{fig1}.
Manifolds with $v+iw=0$ everywhere form the particular class
of manifolds with $SU(3)$ structure.  Those with $c=0$ everywhere form
another particular class, the one of manifolds with {\emph{static}}
$SU(2)$ structure. On such manifolds the pure spinors $\Phi_-$ and
$\Phi_+$ have type one and type two everywhere. The Euler
characteristic of any manifold with static $SU(2)$ structure is zero,
because otherwise the vector field corresponding to $v+iw$ would
have zeroes.

In summary the set of manifolds with $SU(3)\times SU(3)$ structure has two important subclasses:\\


\begin{equation}
\boxed{
\ba
&SU(3) \times SU(3) \hbox{ structure}:\cr
&\qquad\quad   (v+ i w , c)\in S^3
\ea
}
\begin{aligned}
\quad 
\begin{rotate}{30}
$\supset$
\end{rotate}
\qquad 
&
\boxed{
\begin{aligned}
& SU(3) \hbox{ structure}:  \cr
& v+ i w =0,\   |c|=1  
\end{aligned} 
}
\cr
&\cr
\quad
\begin{rotate}{330}
$\supset$
\end{rotate}
\qquad 
&
\boxed{
\begin{aligned}
& \hbox{static }SU(2) \hbox{ structure}:  \cr
& |v+ iw|=1 ,\ c =0
\end{aligned}
}
\end{aligned}
\end{equation}

\vspace{.2cm}


\subsection{Generalized geometry, generalized submanifolds and D-branes}

For the sake of completeness, let us recall a few definitions from generalized complex (GC) geometry
\cite{Gualtieri:2003dx}.  Given an $n$-dimensional manifold $M$, with even $n$, a
generalized almost complex structure on $M$ is defined as an almost complex
structure on the sum of tangent and cotangent bundles $TM\oplus T^\ast M$. 
For example, such a structure can be induced by an ordinary
complex structure $J$ on $M$
\begin{equation}\label{genB}\mathcal{J}_J=\begin{pmatrix}
J  & 0 \\
0 & -J^\ast \hfill 
\end{pmatrix}\,,\end{equation}
in which case it will sometimes be termed a diagonal GC structure, or by a symplectic form $\omega$ on $M$ 
\begin{equation}\label{genA}\mathcal{J}_\omega=\begin{pmatrix}
0  & -\omega^{-1} \\
\omega & 0 \hfill  
\end{pmatrix},\end{equation}
where the matrices are written in a basis adapted to the direct sum
$TM\oplus T^\ast M$.  Hybrid examples, other than these two extreme
ones, are classified by a generalized Darboux theorem
\cite{Gualtieri:2003dx}, saying that any GC space is locally the sum
of a complex space and a symplectic space.  Hybrid GC structures with no 
underlying complex or symplectic
structure do appear in $\mathcal{N}=1$ supersymmetric
compactifications of string theory \cite{Cavalcanti2005,
Grana:2006kf}.

 Around every point  $p$, the sum $T_p M\oplus T_p^\ast M$ is naturally endowed with an inner
 product of signature $(n,n)$,
\be
\langle X+\xi , Y+\eta\rangle=\frac{1}{2}(\iota_X\eta+\iota_Y\xi) \,.
\ee
It also acts naturally on polyforms on $M$:
\be(X+\xi).\phi=\iota_X \phi +\xi\wedge\phi.\ee
Acting twice on $\phi$ yields a Clifford algebra, and the $+i$ eigenbundle of a GC structure is an $n$-dimensional subspace, hence the one-to-one correspondence between GC structures and pure spinors (polyforms with an $n$-dimensional annihilator). On a Calabi--Yau manifold, the pure spinor associated to the diagonal GC structure induced by the ordinary complex structure is the holomorphic $n$-form, while the pure spinor associated to the off-diagonal GC structure induced by the symplectic structure is $e^{i\omega}$, where $\omega$ denotes the K\"ahler form.

The inner product is conserved by an action of the group $O(n,n)$, whose generic element contains
  off-diagonal blocks that can be exponentiated into the so-called $B$- and $\beta$-transforms
\be
\ba
\exp B =\begin{pmatrix}
      1  & 0 \\
      B & 1 \hfill 
\end{pmatrix}\,: \qquad &    B : X+\xi\mapsto X+\xi+\iota_X B \cr
\exp \beta=\begin{pmatrix}
1  & \beta \\
0 & 1 \hfill 
\end{pmatrix}\,:\qquad & \beta : X+\xi\mapsto X+\iota_\xi \beta +\xi \,,
\ea
\ee
where $B$ and $\beta$ are antisymmetric blocks identified with a two-form
$B_{\mu\nu}$ and a bivector $\beta^{\mu\nu}$. The correponding transforms act by conjugation on the matrices of the GC structures, and by left-multiplication by $e^B$ or $e^\beta$ on the corresponding pure spinors. These actions will occur in section 6.

Let $H$ be a closed three-form. A generalized submanifold is defined
in  \cite{Gualtieri:2003dx} as a submanifold $N$ endowed with a two-form $B$ such
that $H|_N=dB$. The generalized tangent bundle $\tau_N^B$ of this
generalized submanifold is defined as the $B$-transform of the sum of
the tangent bundle $TN$  and conormal bundle (or annihilator) $\mathrm{Ann}\,TN$, namely:
\be
\label{GenTanB}
\tau_N^B=\left\{ X+\xi \in TN\oplus T^\ast M |_N, \;\xi|_N=\iota_X B
\right\}\,,
\ee
so that $\tau_N^0= TN\oplus \mathrm{Ann}\,TN$. A generalized tangent
bundle is a maximally isotropic subspace (\ie, it is isotropic with
respect to natural pairing and it has the maximal possible dimension for
an isotropic space in ambient signature $(n,n)$, namely $n$.)
Moreover, all the maximally isotropic subspaces are of this form, for some submanifold $N$ and two-form $B$.

Given a GC structure $\mathcal{J}$, a generalized complex brane is
defined in \cite{Gualtieri:2003dx} as a generalized submanifold
whose generalized tangent bundle is stable under the action of
$\mathcal{J}$. In the case of a diagonal GC structure, the
compatibility condition gives rise to the B-branes, as expected due to
the localization properties of the B-model on complex parameters
\cite{Ooguri:1996ck}.  The submanifold $N$ namely has to be a complex
submanifold, and $F$ has to be of type $(1,1)$ with respect to $J$ 
\be 
J(TN)\subset TN \,,\qquad
 J^\ast(\iota_X F)+\iota_{JX} F =0 \,.
\ee
In the other extreme case of a symplectic structure, the definition
yields all possible types of A-branes, including the non-Lagrangian
ones 
\cite{Kapustin:2001ij,Chiantese:2004pe}. These are two tests of the idea that D-branes
in generalized geometries are generalized submanifolds. This idea has
passed further tests: calibrating forms and pure spinors encoding stability conditions for topological branes \cite{Marino:1999af} are
correctly exchanged by mirror symmetry
\cite{Leung:2000zv,Grange:2004ah, BB1, Koerber:2005qi}.


\section{The SYZ argument for Calabi--Yau manifolds}

Let us sketch the SYZ argument \cite{Strominger:1996it}, assuming for a moment that $M$ is an
ordinary Calabi--Yau manifold with a Calabi--Yau mirror $\hat{M}$. We break the argument up into steps, which we shall then extend to generalized Calabi--Yau manifolds.\\

\noindent
 {\it Step 1: Consider the D0-branes of the B-model on $M$.}\\
 As there is an ordinary complex
structure on $M$, one can always put stable D0-branes on it. In other
words, the moduli space of a D0-brane consists of the entire manifold
$M$.\\

\noindent
 {\it{Step 2: Consider the A-model on the mirror manifold $\hat{M}$.}}\\
 As mirror symmetry does not change moduli spaces, there must be a stable D-brane
$L$ on $\hat{M}$ (a special Lagrangian submanifold (SLag) of
$\hat{M}$) that has the same moduli space, namely $M$. It is safe to disregard the coisotropic D-branes of the A-model in this context \cite{Kapustin:2001ij,Kapustin:2000aa,Font:2006na}, because they are five-dimensional and one eventually considers D-branes that can be obtained from D0-branes by three T-dualities, which rules out dimension five.\\

\noindent
 {\it{Step 3: Project out the gauge-bundle moduli.}}\\
Moreover, this
moduli space has a fibered structure: it is fibered over the set of
geometric moduli called $\mathcal{M}_{\mathrm{SLag}}(L)$, with fiber
given by the gauge-bundle moduli (the projection map $\pi$ is given by
``forgetting the bundle data''):
\begin{equation}\label{fibsyz} M \xrightarrow[]{\pi}
 \mathcal{M}_{\mathrm{SLag}}(L).
\end{equation}
 $M$ is therefore fibered by the gauge bundle data, with fiber given by the set of Wilson lines $T^{b_1(L)}$.\\
 
\noindent
{\it{Step 4: Describe the local tangent space to the moduli space of supersymmetric three-cycles.}} \\
The
 tangent space at $L$ to the moduli space of SLags  \cite{mclean96deformation} with flat
 connections is given  by 
\begin{equation} H^1(L,\mathbf{C})\simeq
 H^1(L)\oplus H^1(L),
\end{equation}
 with the first term corresponding to geometric
 moduli and the second one to gauge-bundle moduli (the
 Lagrangian and special condition are preserved by exactly those
 deformations that are induced by harmonic one-forms, and the flat
 gauge connections are described by the set of $b_1$ monodromies
 around the non-trivial homology cycles in $L$).\\

\noindent
{\it{Step 5: Use the result of step 1 to compute the dimension of the fibers.}} \\
The moduli space of
 SLags with flat connections on $\hat{M}$ (continuously connected to
 $L$) therefore has real dimension $2b_1(L)$, half of which comes from
 the moduli of flat connections. But the fiber in the fibration
 (\ref{fibsyz}) is a torus $T^{b_1(L)}$. As this moduli space is $M$
 itself, we learn that $2b_1=6$, and that $M$ is fibered by
 three-tori.\\

\noindent
 {\it{Step 6: T-dualize along the three-cycles.}} \\
 Consider a D3-brane with flat connection wrapping a $T^3$ fiber on
  $M$. T-dualizing along the three $U(1)$ directions produces a
  D0-brane on a T-dual manifold called $M'$, whose moduli space is the
  whole of $M'$.  Consider a D0-brane on $M$. Its moduli space is the
  whole of $M$. It sits at some point in a $T^3$ fiber. T-dualizing
  along the three $U(1)$ directions of this fiber produces a D3-brane
  with flat connection wrapping a three-cycle on $M'$. This describes
  a fibration of $M'$ by three-tori, whose moduli space is $M$. This
  is the same situation as with the couple of branes on $M$ and
  $\hat{M}$ described above. Therefore $\hat{M}=M'$ and T-duality
  along the torus fibers is mirror symmetry.


\section[Fibrations {\emph{\`a la}} SYZ for static $SU(2)$ structure manifolds]{Fibrations {\emph{\`a la}} SYZ for static $SU(2)$ structure \\ manifolds}

 Manifolds with static $SU(2)$ structure form an interesting but still
 tractable subclass of backgrounds because they substantially differ
 from Calabi--Yau manifolds (in that they admit no closed type-three
 pure spinor), and because they do not exhibit type-jumping
 phenomena. They are relatively tractable, for the price of
 considering cycles that are transverse to $M$ and its mirror
 $\hat{M}$. Having type-one and type-two closed pure spinors, we find
 it natural to form their wedge product, which induces a GC structure
 on the product $M\times \hat{M}$, because the wedge product starts
 with a complex three-form and allows for some parallel treatment of
 the SYZ argument.


\subsection{Supersymmetric cycles on $M\times \hat{M}$}

Consider a generalized Calabi--Yau manifold $M$ and its mirror
$\hat{M}$, both with static $SU(2)$-structure.\footnote{For recent developments based on the physics of $SU(2)$ structure manifolds as gravity duals of deformations of super Yang--Mills theories, see for instance \cite{Minasian:2006hv}.} 
There is a nowhere-vanishing complex one-form field $v+iw$, inducing on every
local four-dimensional transverse space a real two-form $j$ and a
complex two-form $\omega$. The corresponding two pure spinors are
\be
\ba
\Phi_-  &=(v+iw)\wedge e^{i j}\cr
\Phi_+ &=\omega \wedge e^{i v\wedge w} \,.
\ea
\ee
They are exchanged under mirror symmetry
with analogous objects on the mirror $\hat{M}$ built from a
nowhere-vanishing complex one-form field $\hat{v}+i\hat{w}$, inducing
on every local four-dimensional transverse space a real two-form
${\hat{j}}$ and a complex two-form $\hat{\omega}$:
\be
\ba
\hat{\Phi}_- &=(\hat{v}+i\hat{w})\wedge e^{i\hat{j}}\cr
\hat{\Phi}_+ &=\hat{\omega}\wedge e^{i \hatv\wedge \hatw} \,.
\ea
\ee
This is a case of the generalized Darboux theorem with types one and two, and we
can choose local coordinates that are adapted to it: 
\be
\ba
j=:dx^3\wedge dx^4+ dx^5\wedge dx^6\,,\qquad &
\hat{j}=:d\hat{x}^3\wedge d\hat{x}^4+d\hat{x}^5\wedge d\hat{x}^6,\cr
\omega=: d(x^3+ ix^4)\wedge d(x^5+ix^6) \,, \qquad  &
\hat{\omega}=:d(\hat{x}^3+ i\hat{x}^4)\wedge d(\hat{x}^5+i\hat{x}^6) \,.
\ea
\ee
The pure spinors $\Phi_-$, $\Phi_+$,  $\hat{\Phi}_-$ and $\hat{\Phi}_+$ induce almost 
GC structures on $M$ and $\hat{M}$ denoted by ${\mathcal{J}}_-$,
${\mathcal{J}}_+$, $\hat{\mathcal{J}}_-$ and $\hat{\mathcal{J}}_+$.\\

\noindent
{\it Step 1.} Where can we place points? 
In order to parallel the first step of the SYZ argument for
 Calabi--Yau manifolds, we need to be able to move points on a
 six-dimensional space. This cannot be $M$ or $\hat{M}$, because
 the GC structure induced by $\Phi_-$ always maps some tangent vectors
 to some normal vectors. This prevents the generalized tangent bundle
 to a point from being stable under the action of the GC structure.

Consider instead supersymmetric cycles on $M\times \hat{M}$. There are
several possible choices for structures and calibrations, and we will
be interested in the following combinations:
\begin{itemize}
\item  GC branes w.r.t. the GC structure ${\mathcal{J}}_- \oplus \hat{\mathcal{J}}_+$, calibrated by ${\Phi}_+ \wedge \hat{\Phi}_-$, which we call $\Sigma$
\item  GC branes w.r.t. the GC structure ${\mathcal{J}}_+ \oplus \hat{\mathcal{J}}_-$,   calibrated by ${\Phi}_- \wedge
\hat{\Phi}_+$,  which we call $\hat{\Sigma}$.
 \end{itemize}
 
In a basis of the local tangent space to $M\times \hat{M}$ adapted to
the local splitting into $2+4$ dimensions, we have the following matrix representation for the GC
structures, where the symbols $J_\omega$ and $J_{\hat{\omega}}$ denote the almost complex structures corresponding to 
$\omega$ and $\hat{\omega}$ in the local four-dimensional subspaces, so that we obtain 
\begin{equation}
\ba 
\mathcal{J}_-\oplus \hat{\mathcal{J}}_+
& 
=\left(\begin{array}{cccc}
 J_{v+iw} & 0 & 0 & 0\\
0 & -J_{v+iw}^\ast & 0 & 0\\
0 & 0 & 0 & -j^{-1}\\
 0 & 0 & j & 0\\ 
\end{array} \right)
\oplus
\left(\begin{array}{cccc}
 0 & -{\hatv\wedge\hatw}^{-1} & 0 & 0\\
 {\hatv\wedge\hatw} & 0 & 0 & 0\\
 0 & 0 & J_{\hat{\omega}} & 0 \\
 0 & 0 & 0 & -J^\ast_{\hat{\omega}}\\
\end{array}\right),
\cr
\mathcal{J}_+\oplus \hat{\mathcal{J}}_-
& 
=\left( \begin{array}{cccc}
0 & -v\wedge w^{-1} & 0 & 0\\
v\wedge w & 0 & 0 & 0\\
0 & 0 & J_\omega & 0\\
 0 & 0 & 0 & -J^\ast_\omega\\
\end{array}\right)
\oplus
\left(\begin{array}{cccc}
J_{\hatv+i\hatw} & 0 & 0 & 0\\
0 & -J_{\hatv+i\hatw}^\ast & 0 & 0\\
0 & 0 & 0 & -\hat{j}^{-1}\\
 0 & 0 & \hat{j} & 0\\
\end{array}\right).
\ea
\end{equation}

Let us describe the generalized tangent bundle $\tau_\Sigma^0$ (with
zero field strength), of the GC
submanifold $\Sigma$ of $M\times \hat{M}$. As the two GC structures we
consider on $M\times \hatM$ are block-diagonal with blocks of the same
size , the projections of the generalized tangent bundle onto the sums
of blocks and dual blocks are separately generalized complex and
calibrated w.r.t. the corresponding blocks.

\begin{figure}[th]
\centerline{\includegraphics[width=7cm]{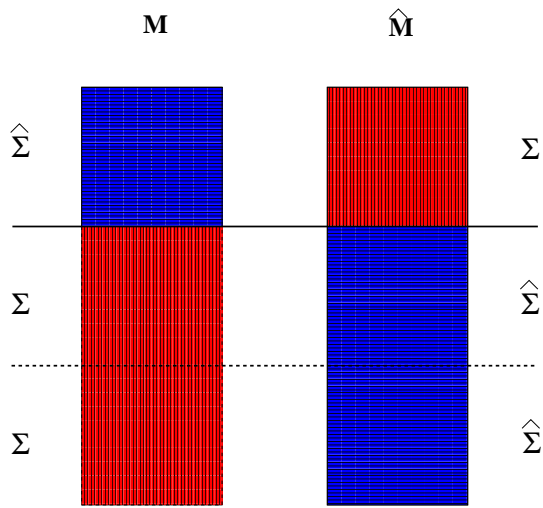}}
\caption{\label{fig2}  
Supersymmetric cycles $\Sigma$ and $\hat{\Sigma}$ and their location within $M \times \hat{M}$.
}
\end{figure}

We may  choose $\Sigma$ to have zero-dimensional projections onto $\vect(v,w)$ and
$\vect(\hatv,\hatw)^\perp$.
chosen to be trivial, the projections of $\tau^0_\Sigma$ onto
$\vect(\hatv,\hatw)$ and $\vect(v,w)^\perp$ have to be Lagrangian
w.r.t. $\hatv\wedge\hatw$ and $j$ respectively, and calibrated by
$\hat{v}+i\hat{w}$ and $\omega$. This gives one-dimensional and two-dimensional
projections on $\vect(\hatv,\hatw)$ and $\vect(v,w)^\perp$
respectively for the world-volume of $\Sigma$
\begin{eqnarray}
j+ \hat{v}\wedge\hat{w}|_\Sigma                                                 &=&0 \label{genLag} \\ 
{\mathrm{Im}}\left(\omega\wedge(\hat{v}+i\hat{w})\right)|_\Sigma  &=&0\,. \label{genSpec}
\end{eqnarray}
They look like the Lagrangian and special conditions, but live on a six-dimensional subspace of $M\times {\hat{M}}$, transverse to both $M$ and $\hat{M}$.
To sum up, a possible local generalized tangent bundle is given in the coordinates chosen above
as:
\be
\tau_{{\Sigma}}^0=\left\langle v, w, \frac{\p}{\p x^3}+ i dx^4,
\frac{\p}{\p x^5} +i dx^6, \hatv_\ast+ i \hatw_\ast, d\hatx^3, d\hatx^4,
d\hatx^5, d\hatx^6 \right\rangle\,.
\ee
The supersymmetric cycle $\Sigma$ is therefore three-dimensional, but
neither of its projections on $M$ or $\hat{M}$ is (they are two- and
one-dimensional respectively). The situation is depicted in fig. \ref{fig2}.

 The same linear-algebraic exercise can be repeated with hats
 exchanged to yield the local generalized tangent bundle of the cycle
 called $\hat{\Sigma}$ (with a somewhat misleading notation because
 $\hat{\Sigma}$ is not mirror to $\Sigma$; both are their own mirror):
\begin{eqnarray}
 {v}\wedge{w} +\hat{j}|_{\hat{\Sigma}}                                               &=&0 \label{hatgenLag}  \\
{\mathrm{Im}}\left(({v}+i{w})\wedge \hat{\omega}\right)|_{\hat{\Sigma}} &=&0 \,,\label{hatgenSpec}
\end{eqnarray}
 with the generalized tangent bundle given by
\be
\tau_{{\hatSigma}}^0=\left\langle v_\ast+ i w_\ast , dx^3, dx^4,
dx^5, dx^6,  \hatv, \hatw, \frac{\p}{\p \hatx^3}+ i
d\hatx^4, \frac{\p}{\p \hatx^5} +i d\hatx^6 \right\rangle \,.
\ee
We thus obtain the situation in Figure \ref{fig2}. The supersymmetric
cycles we have just described are sketched as submanifolds of $M\times
\hat{M}$ that are transverse to both $M$ and $\hat{M}$, whereas the
tree-dimensional supersymmetric cycles on a mirror pair of Calabi--Yau
manifolds are longitudinal either to $M$ or to $\hat{M}$. If one
thinks of a supersymmetric three-cycle as a leg, then the SYZ picture
of mirror pairs correspond to standing on $M\times\hat{M}$ with one
leg on $M$ and one leg on $\hat{M}$ . What we have just argued is that standing on $M\times \hat{M}$ with $M$ and $\hat{M}$ generalized Calabi--Yau manifolds with static $SU(2)$ structure can be achieved, but only with the legs crossed.


\subsection{The fibration of $M\times\hat{M}$}

We are ready to turn to {\it Step 3} and {\it Step 4}.
So far we have exhibited two three-dimensional supersymmetric cycles
on $M\times\hat{M}$, called ${\Sigma}$ and ${\hat{\Sigma}}$, each of which possesses six position moduli given by the projections onto the subspaces of $M\times \hat{M}$ that are complex w.r.t. $(v+iw)\wedge\hat{\omega}$ and $\omega\wedge(\hat{v}+i\hat{w})$. We want
to show that they both have the topology of a three-torus, and that
the moduli space of $\Sigma\times \hat{\Sigma}$ on $M\times\hat{M}$ is
the whole of $M\times\hat{M}\times M\times\hat{M}$.

 We have worked out the generalized tangent spaces of both cycles
 ${\Sigma}$ and ${\hat{\Sigma}}$. This gives only local informations,
 essentially counting dimensions. For some point $p$ in
 $M\times\hat{M}$, on the local tangent space $T_p(M\times\hat{M})$ in
 which $\Phi_-$, $\Phi_+$, $\hat{\Phi}_-$ and $\hat{\Phi}_+$ have the
 generalized Darboux expressions we wrote above, the projections of
 $\tau_\Sigma^0$ onto the subspace $\vect(v,w)$ and
 $\vect(\hat{v},\hat{w})^\perp$ have dimension zero. So have the
 projections of $\tau_{\hat{\Sigma}}^0$ on $\vect(v,w)^\perp$ and
 $\vect(\hat{v},\hat{w})$. We have just described a projector
 \be
 \tau_\Sigma^0\times\tau_{\hat{\Sigma}}^0 \rightarrow
 T_p(M\times\hat{M})\,.
 \ee
This is the tangent application to the
 projection 
 \be
\Sigma \times \hat{\Sigma}  \mapsto  p \in  M\times\hat{M} \,.
 \ee 
 There are therefore position moduli for  $\Sigma\times\hat{\Sigma}$ in all the twelve directions, which
 correspond to moving $\Sigma\times \hat{\Sigma}$ around $p$ in the
 coordinate patch.

 Let us call $\mathcal{M}$ the moduli space of supersymmetric cycles
 of $M\times\hat{M}$ that are continuously connected to
 $\Sigma\times\hat{\Sigma}$. We have just argued that there is a
 twelve-dimensional subspace consisting of translation moduli, so there must exist
 other moduli, which make up some subspace $\mathcal{M}'$
 consisting of deformations that leave the projection of
 $\Sigma\times\hat{\Sigma}$ onto the local complex subspaces of
 $M\times \hat{M}$ fixed: 
 \be
 T_{\Sigma\times
 \hat{\Sigma}}\mathcal{M}=T_p(M\times\hat{M})\times T_{\Sigma\times
 \hat{\Sigma}}\mathcal{M}' \,.
 \ee

Consider now the projection $\pi$ that ``forgets the gauge bundle'' along the two cycles $\Sigma$ and $\hat{\Sigma}$. It induces a fibration of $\mathcal{M}'$ over some base $\mathcal{B}$ consisting of the Lagrangian deformations of $\Sigma\times\hat{\Sigma}$ w.r.t. $v\wedge w\wedge j\wedge\hat{v}\wedge\hat{w}\wedge \hat{j}$:
\be
\label{genfib}\mathcal{M}'\xrightarrow[]{\pi}
 \mathcal{B}
\ee
The space $\mathcal{M}'$ has the following topological
 meaning, as the fiberwise projection of the generalized tangent
 bundle is isomorphic to the complexified dual of the ordinary tangent
 space to $\Sigma\times\hat{\Sigma}$ as a bundle and as a Lie
 algebroid (cf. section 7.2 of \cite{Koerber:2006hh}) :
\be
{\mathcal{M}}'=H^1\left(L(\Sigma\times\hat{\Sigma},0)|_p\right)
=H^1_{\mathrm{dR}}(\Sigma\times\hat{\Sigma},\mathbf{C}).
\ee
 The dimension of the moduli space $\mathcal{M}'$ is therefore twice the
 Betti number of the six-dimensional cycle $\Sigma\times
 \hat{\Sigma}$.

 In order to compute this dimension, we are going to perform T-duality along
 $\Sigma\times\hat{\Sigma}$, with the point $p$ still fixed. This
 implies that gauge connections will be fixed on the image of
 $\Sigma\times\hat{\Sigma}$, which will only have translation
 moduli. The moduli space is not changed by T-duality, but it is now
 described as follows. The one-dimensional projection of $\Sigma$ onto
 $T_pM$, with moduli from flat connection and normal deformations (all in
 $\mathcal{M}'$), is mapped to a point-like projection onto $T_p\hat{M}$
 with two translation moduli. The fixed zero-dimensional projection
 of $\Sigma$ onto $T_p\hat{M}$, is mapped to a one-dimensional projection
 onto $T_p M$ sitting at some fixed $w_\ast=0$, extended along $v_\ast$
 and with fixed flat connection. The two-dimensional projection of
 $\Sigma$ onto $T_p\hat{M}$ is mapped to a point of $M$ with four
 translation moduli. So $\Sigma$ is mapped to itself,
 but the fiber moduli are traded for translational ones, living in the
 subspace $\vect(\hat{v_\ast},\hat{w_\ast})\oplus
 \vect(v_\ast,w_\ast)^\perp$. By T-dualizing $\hat{\Sigma}$ along
 $\Sigma\times\hat{\Sigma}$, one trades in the same way the
 deformation moduli for translational ones, living in the subspace
 $\vect({v_\ast},{w_\ast})\oplus
 \vect(\hat{v_\ast},\hat{w_\ast})^\perp$, so that the tangent space at
 $\Sigma\times \hat{\Sigma}$ to the moduli space $\mathcal{M}'$ is
 isomorphic (by dimension counting) to 
\be
\vect(\hat{v_\ast},\hat{w_\ast})\oplus
 \vect(v_\ast,w_\ast)^\perp\oplus \vect({v_\ast},{w_\ast})\oplus
 \vect(\hat{v_\ast},\hat{w_\ast})^\perp=  T_p (M\times\hat{M})\,.
 \ee
 The moduli (sub)-space $\mathcal{M}'$ therefore has dimension
 twelve.

Let us move to {\it Step 5}.
We have just computed the dimension of $\mathcal{M}'$, which is
 accessible to our local computations, but its T-dual interpretation
 in homology promotes the result to a Betti number, a global
 quantity. From this T-duality argument we learn that 
 \be
 \mathrm{dim}
 T_{\Sigma\times \hat{\Sigma}}\mathcal{M}'= 2 b_1(\Sigma \times
 \hat{\Sigma})=\mathrm{dim} (M\times \hat{M})=12 \,.
 \ee
 Hence $\Sigma
 \times \hat{\Sigma}$ is a six-torus, the product of two
 supersymmetric three-tori, and its moduli space is
 $M\times\hat{M}\times M\times \hat{M}$.

Note that $\Sigma$ or $\hat{\Sigma}$ by itself does not have $M\times \hat{M}$ as its moduli space, nor $M\times M$ nor $\hat{M}\times\hat{M}$, as it is only the case for $SU(3)$ structures.


\section{Illustrations in flux compactifications}

 So far we have drawn the conclusions of there being transverse three-dimensional supersymmetric cycles on a mirror pair of manifolds with static $SU(2)$ structures. This begs for a few checks. We shall first T-dualize the three-tori and check that the pure spinors are exchanged by this transformation. We shall then turn to the example of $K3\times T^2$, which was of course available in the Calabi--Yau case, but can also be endowed with a static $SU(2)$ structure. Finally, in order to make contact with open problems in flux compactifications (where the nature of non-geometric fluxes is still under investigation), we shall take the analog of {\it Step 6} by turning on all the possible fluxes on a six-torus with static $SU(2)$-structure, thus putting our T-duality proposal to the test.

\subsection{Mirror images of the pure spinors}

Let us perform a Fourier--Mukai transform ($\ft$) on the pure spinors, by
weighting them with the Poincar\'e connection on $\Sigma\times
\hat{\Sigma}$ we worked out. As we have established that the
three-dimensional intersection of $\Sigma\times \hat{\Sigma}$ and $M$
 are the directions which are T-dualized, the Fourier--Mukai transform of the pure spinors reads   
\be
\ft\left({\Phi_-}\right)=\int_{(\Sigma\times\hat{\Sigma})\cap M}
(v+iw)\wedge e^{ij}\wedge e^{v\wedge \hat{v} +dx^3\wedge d\hatx^3 +
dx^5\wedge d\hatx^5}= e^{i\hat{v}\wedge w} \wedge \omega=
\hat{\Phi}_+,\ee 
\be\ft\left({\Phi_+}\right)=\int_{(\Sigma\times\hat{\Sigma})\cap M}
e^{iv\wedge w}\wedge \omega\wedge e^{v\wedge \hat{v} +dx^3\wedge
d\hatx^3 + dx^5\wedge d\hatx^5}= (\hat{v}+iw) \wedge e^{i\hat{j}}=
\hat{\Phi}_-,\ee 
 with the value of the base coordinates unchanged,
namely provided $w=\hat{w}$, which makes sense, because the local
coordinates $w$ or $\hat{w}$ are not T-dualized. The mapping of pure spinors under Fourier--Mukai transform coincides with what is expected from mirror symmetry.


\subsection{The  $K3\times T^2$ example}

As the Euler characteristic is multiplicative, the manifold $K3\times
T^2$ has Euler characteristic zero. There may therefore be a
nowhere-vanishing vector field on it. Real and imaginary part of the
complex coordinate of $T^2$ as an elliptic curve indeed serve as $v$ and $w$
vector fields.\footnote{For a thorough treatment of the reduction of IIA supergravity on $K3\times T^2$ endowed with an $SU(2)$ structure, see \cite{bastiaanJan}.}

  In the present case, $\omega$ and $j$ are a complex and a K\"ahler form on $K3$, while
  $\hat{\omega}$ and $\hat{j}$ are the same objects on the mirror
  $K3$. Of course in this case we have a global picture of the cycles:
  $\Sigma$ is a point in $T^2$ times a special Lagrangian torus with
  respect to $j$, times a Lagrangian circle in the mirror torus times
  a point in the mirror $K3$, while $\hat{\Sigma}$ is the mirror
  circle on the first torus times a point in the first $K3$ times a
  point in the second $T^2$ times the dual torus in the second
  $K3$. The projection is just given by associating the points to
  $\Sigma\times\hat{\Sigma}$. This is just the ordinary SYZ case but
  with the complex structures of the two-tori exchanged. It is a
  straightforward consequence of the Calabi--Yau case because crossing the legs amounts to permuting the two two-tori.


\subsection{Static $SU(2)$ structure with non-geometric fluxes}

 Let us apply this analysis to the case of a six-torus endowed with a static $SU(2)$ structure. 
 This seems of course to be an over-simplification, as many torus fibrations can be explicitly found in such a geometry. 
 However, T-duality leads from geometric to non-geometric fluxes, which in the terminology \cite{Shelton:2005cf} are called $Q$- and $R$-fluxes according to the number of T-dualized directions supporting a $B$-field. With each double arrow symbolizing one T-duality, these notations are summarized in the following way: 
\be
H_{abc}\leftrightarrow  f^a_{bc}\leftrightarrow Q^{ab}_c \leftrightarrow
R^{abc}.
\ee
 The embedding of three-tori into 
 $M\times\hat{M}$ along which T-duality is performed is key to the map between geometric and non-geometric fluxes. 
 Finding the mirror of a generic flux configuration is therefore a non-trivial check of our 
 proposal\footnote{Choosing a static $SU(2)$ structure protects us against type-jumping phenomena; 
 those will of course be crucial in the generic $SU(3)\times SU(3)$ case, which will be elaborated on in the next section, in a much less thorough way though.}.   
 We are going to complete the study of fluxes on the $SU(2)$ structure background of  
 \cite{Grana:2006hr}, first including all the non-geometric fluxes 
 (which indeed fill all the entries of the charge matrix), and then to obtain the mirror configuration by T-duality along the transverse supersymmetric fibers.  


\subsubsection{Charge matrix}

We consider a six-torus endowed with a static $SU(2)$ structure.
The holomorphic vector $e^3= v+iw$ is completed to a basis by $(e^1,e^2,e^3)$, and likewise for the mirror the basis is denoted by $(\hat{e}^1, \hat{e}^2, \hat{e}^3)$.
The GC submanifolds $\Sigma$ and $\hat\Sigma$ solving the structure and stability equations (\ref{genLag})-(\ref{hatgenSpec}) are chosen as
\be
\begin{array}{l|l}
\Sigma & \hat\Sigma\cr
\hline
\Re(e^1) & \Re(\hat{e}^1)\cr\hline
\Re(e^2) & \Re(\hat{e}^2)\cr\hline
\Re(\hat{e}^3) & \Re(e^3)
\end{array} \,,
\ee
which have trivial projection onto the base spanned by $(\Im(e^1), \Im(e^2), \Im (e^3))$ and $\Sigma$ projects trivially upon $e^3= v+iw$ etc. as required.

The generic $SU(3)\times SU(3)$ structure
is described by a symplectic basis with forms that are not necessarily closed. 
Denote the two bases by
\be
\Sigma^- = \left( 
 \begin{array}{c}
 \alpha_I  \cr
 \beta^I
 \end{array}
 \right)\,,\qquad 
 \Sigma^+ = \left( 
 \begin{array}{c}
 \omega_A  \cr
 \tilde{\omega}^B
 \end{array}
 \right) \,,
\ee
where the entries of $\Sigma^{\pm}$ are odd/even formal sums of forms. In particular $d\Sigma^{\pm} \not =0$ and can therefore be expanded in $\Sigma^{\mp}$, {\it  i.e.}
\be \label{QDef}
d \Sigma^- = \mathcal{Q} \Sigma^+ \,.
\ee
The matrix $\mathcal{Q}$ is called the charge matrix. In the present case it is a four-by-four matrix.

Furthermore define the generalized symplectic basis $\Sigma^{\pm}$ in terms of the basis $e^i$ as follows
\begin{equation}
\Sigma^- =
\left(
 \begin{array}{c}
  2 \Re (e^3) \\
  -2 \Im (e^3) +  \Re (e^3) j \\
  - \Im (e^3 ) j^2 \\
  {1\over 3} \Re (e^3) j^2 + {4\over 3} \Im (e^3) j
 \end{array}
\right),
\end{equation}
and
\begin{equation}
\begin{aligned}
\Sigma^+ 
& =
\left(
 \begin{array}{c}
  4  \Re ( e^1)  \wedge \Re (e^2) \\
  8 \left( \Im (e^1) \wedge \Re(e^2) + \Re (e^1) \wedge \Im (e^2) \right) 
  - 16 \Re (e^1) \wedge \Re(e^2) \wedge \Re( e^3) \wedge \Im(e^3)\\
   16 \Im (e^1) \wedge \Im(e^2) \wedge \Re(e^{3}) \wedge \Im (e^3) \\
   {4\over 3} \Im (e^1) \wedge \Im (e^2) 
   + {4\over 3}  (\Im (e^1) \wedge \Re(e^2) + \Re(e_1) \wedge \Im (e^2)) \wedge
     \Re (e^{3}) \wedge \Im (e^3)
 \end{array}
\right)\,,
\end{aligned}
\end{equation}
where we defined 
\begin{equation}
j = 2 i (e^1 \wedge e^{\bar{1}} + e^2 \wedge e^{\bar{2}}) 
  = 4 \left(\Re (e^1) \wedge \Im (e^1) +\Re (e^2) \wedge \Im (e^2)  \right)
\,.
\end{equation}

 As discussed earlier, the standard relation between the two symplectic basis vectors is (\ref{QDef}).
Turning on fluxes -- both geometric $H$-flux and non-geometric $Q$- and $R$-fluxes -- has the effect of twisting the the differential operator $d$ 
\begin{equation} 
(d + H\wedge + \,Q \cdot + \,R \cdot) \Sigma^- \sim  \mathcal{Q}\Sigma^+ \,.
\end{equation}
Here we denote by $\sim$ equality up to terms that are perpendicular to all elements in the symplectic basis with respect to the symplectic pairing 
\be
\int_M \langle \sigma, \rho \rangle = \int_M \left(\sum_{p} (-1)^{[{p+1\over 2}]}  \sigma_p \wedge \rho_{6-p}\right) \,,
\ee
where $\sigma= \sum_p \sigma_p$ is a polyform (the sum runs over the degrees) 
and $\langle, \rangle$ denotes the Mukai pairing.
In particular the symplectic basis obeys
\be
\int_M \langle \alpha^I, \beta_J \rangle = \delta_I^J \,,\qquad 
\int_M \langle \omega_A, \tilde{\omega}^B \rangle= \delta_A^B \,.
\ee

\begin{figure}[th]
\centerline{\includegraphics[width=8cm]{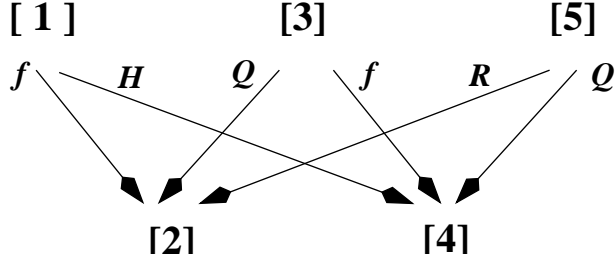}}
\caption{\label{fig3} \small Mapping of cohomology degrees under the fluxes $f$, $H$, $Q$ and $R$.}
\end{figure}

Note that the action on cohomologies is as follows
\begin{equation}
\begin{aligned}
d &:\qquad H^p \rightarrow H^{p+1} \cr
H &:\qquad H^p \rightarrow H^{p+3}  \cr
Q\cdot &: \qquad H^p \rightarrow H^{p-1} \cr
R\cdot &: \qquad H^p \rightarrow H^{p-3} \label{CohoRels}\,,
\end{aligned}
\end{equation}
in agreement with $Q$ having two vector and one form index and $R$ being a tri-vector.  
Note that $d$ acts on the one-forms as $d e^{i} = f^{i}_{jk} e^j \wedge e^k$.
The mapping of the various degrees under the fluxes (\ref{CohoRels}) can be depicted as in Fig. \ref{fig3}. Here $[p]$, with $p= 1,2,\cdots$ denotes the degree of the forms.

The various flux components then follow by noting that 
\begin{equation}
\Sigma^- = \left( \begin{array}{c}  [1] \cr [1] + [3] \cr [5] \cr [3] + [5] \end{array}\right) \,,\qquad 
\Sigma^+ = \left( \begin{array}{c}  [2] \cr [2] + [4] \cr [4] \cr [2] + [4] \end{array}\right) \,,
\end{equation}
and further allowing additional terms compatible with the equivalence relation $\sim$.


\subsubsection{Geometric fluxes}

The effect of the geometric fluxes (both $H$ and $f$) was already discussed in \cite{Grana:2006hr}. 
There it was found that with the geometric flux parameters one can switch on the following entries in the charge matrix
\begin{equation} \label{GeoQ}
(d + H\wedge) \Sigma^-
\sim
\left(
\begin{array}{c|c|c|c}
 \mathbb{F}_{11}  & \mathbb{F}_{12} +\mathbb{H}_{12} &  \mathbb{H}_{13}& \mathbb{F}_{14}+\mathbb{H}_{14} \cr \hline
 \mathbb{F}_{21} & \mathbb{F}_{22} + \mathbb{H}_{22}& \mathbb{F}_{23} + \mathbb{H}_{23}&\mathbb{F}_{24}+ \mathbb{H}_{24} \cr \hline
0&0&0&0 \cr\hline
0 & \mathbb{F}_{42} & \mathbb{F}_{43} &\mathbb{F}_{44}   
\end{array}
\right) 
\Sigma^+ \,.
\end{equation}

The geometric flux charges (a.k.a. torsion charges) $\mathbb{F}_i$ follow from the 
relation
\begin{equation}
d \mathfrak{e}^I = f^{I}_{JK} \mathfrak{e}^{J} \wedge \mathfrak{e}^{K}\,,
\end{equation}
where $\mathfrak{e}^{I} = \Re (e^i)$ for $I=i$ and $\mathfrak{e}^{I}=\Im (e^i)$ for $I= \bar{i}$.

To sum up, 
the $f$-flux we have to turn on in order to generate the above charge entries are
\begin{equation}
\begin{aligned}
f
    =& +2 \mathbb{F}_{11} \Re (\partial_3) \wedge \Re (e^1) \wedge \Re (e^2)\cr
      &+ 8 \mathbb{F}_{12} \left(\Re (\partial_3) \wedge \Im (e^1) \wedge \Re (e^2)  
              + \Re (\partial_3) \wedge \Re (e^1) \wedge \Im (e^2) \right) 
   \cr
   &+ {4\over 3} \mathbb{F}_{14} \Re (\partial_3) \wedge \Im (e^1) \wedge \Im (e^2)  
   \cr
      & - 2 \mathbb{F}_{21} \Im(\partial_3) \wedge \Re (e^1) \wedge \Re (e^2) \cr
     &-4 \mathbb{F}_{22} ( \Im (\partial_3) \wedge \Im(e^1) \wedge \Re (e^2)
                          +\Im (\partial_3) \wedge \Re(e^1) \wedge \Im (e^2) ) \cr 
     & - 2 \mathbb{F}_{22} ( \Im(\partial_1) \wedge \Re (e^2) \wedge \Im(e^3)
                            + \Im(\partial_2) \wedge \Im(e^3) \wedge \Re (e^1)) \cr  
     & + 2 \mathbb{F}_{23} (\Re (\partial_1)\wedge \Im (e^3) \wedge \Im (e^2)
                 + \Re (\partial_2) \wedge\Im(e^1) \wedge \Im (e^3) )\cr
     & - {1\over 3} \mathbb{F}_{24} \left(2 \Im (\partial_3)\wedge \Im(e^1) \wedge \Im(e^2) 
                + \Re (\partial_1)\wedge \Re (e^2) \wedge \Im(e^3)  
                  + \Re (\partial_2) \wedge \Im(e^3) \wedge \Re (e^1) \right)\cr
     & +  3 \mathbb{F}_{42} (- \Re (e^2) \wedge \Im (\partial_1) + \Re (e^1) \wedge \Im (\partial_2)) \wedge \Re (e^3) \cr
     &+ {3\over 2} \mathbb{F}_{43} (\Re (\partial_1) \wedge \Im (e^2) \wedge \Re (e^3) 
           + \Re (\partial_2) \wedge \Re (e^3) \wedge \Im (e^1)) \cr 
        & +{1\over 2} \mathbb{F}_{44} (\Im (e^2) \wedge \Im (\partial_1) + \Im (\partial_2) \wedge \Im (e^1)) \wedge \Re (e^3)  \,.
   \end{aligned}
\end{equation}
We should perhaps add a word of explanation. Recall that the relations between the two symplectic basis is only up to the equivalence w.r.t. $\sim$. This in particular allows one to switch on $f$-flux to generate the $\mathcal{Q}_{12}$ charge entry, without turning on $H$-flux simultaneously. To be more explicit
\be
f=  8 \mathbb{F}_{12} \left(\Re (\partial_3) \wedge \Im (e^1) \wedge \Re (e^2)  
              + \Re (\partial_3) \wedge \Re (e^1) \wedge \Im (e^2) \right) 
\ee
acting upon $\Sigma^-_1 = 2 \Re (e^3)$ will only generate the two-form part of $\Sigma^+_2$, denoted by $\left.\Sigma_2^+\right|_{[2]}$
\be
f \Sigma^-_1 = 2 \mathbb{F}_{12} \left.\Sigma_2^+\right|_{[2]} \,.
\ee
However, this can be written as
\be
f \Sigma^-_1 =  \mathbb{F}_{12} \Sigma_2^+ + \Omega \,,
\ee
where
\be
\Omega =  \mathbb{F}_{12} \left( \left.\Sigma_2^+\right|_{[2]}- \left.\Sigma_2^+\right|_{[4]} \right) \,,
\ee
which is perpendicular to all other basis elements
\be
\langle \Omega, \Sigma^\pm_i \rangle =0 \,,
\ee
and thus 
\be
f \Sigma^-_1 \sim \mathbb{F}_{12} \Sigma_2^+\,.
\ee

Likewise the $H$-flux can be determined as
\begin{equation}
\begin{aligned}
H =&       +  16 \mathbb{H}_{12} \Re (e^1) \wedge \Re (e^2) \wedge \Im (e^3) \cr
       &             -8 \mathbb{H}_{13} \Im (e^1) \wedge \Im (e^2) \wedge \Im (e^3)  \cr
   &  +  {4\over 3} \mathbb{H}_{14} \left(\Re (e^2) \wedge \Im (e^1 )- \Re (e^1) \wedge \Im (e^2) \right) \wedge \Im (e^3)  \cr
   &  +  16 \mathbb{H}_{22} \Re (e^1) \wedge \Re (e^2) \wedge \Re (e^3) \cr
    &    - 8 \mathbb{H}_{23} \Im (e^1) \wedge \Im (e^2) \wedge \Re (e^3) 
    \cr
   &  -  {4\over 3} \mathbb{H}_{24} \left(\Im (e^1 )\wedge \Re (e^2) \wedge \Re (e^3) 
      + \Re (e^1) \wedge \Im (e^2) \wedge \Re (e^3) \right) \,.
\end{aligned}
\end{equation}
The resulting charge matrix entries are as we indicated in (\ref{GeoQ}).


\subsubsection{Non-geometric fluxes}

Here we wish to study the effect of the $Q$- and $R$-fluxes, which can be done by linear superposition with the results from \cite{Grana:2006hr}. 
We find by simple dimensional analysis that the effect of these non-geometric fluxes on the charge matrix can be only of the following type:
\begin{equation}
(Q+R) \Sigma^- \sim
\left(
\begin{array}{c|c|c|c}
0&0&0&0 \cr\hline
\mathbb{Q}_{21} & \mathbb{Q}_{22}&0&\mathbb{Q}_{24} \cr\hline
\mathbb{R}_{31} & \mathbb{Q}_{32} + \mathbb{R}_{32} &\mathbb{Q}_{33}& \mathbb{Q}_{34} + \mathbb{R}_{34} \cr \hline
\mathbb{R}_{41}&\mathbb{Q}_{42}+ \mathbb{R}_{42}&\mathbb{Q}_{43} & \mathbb{Q}_{44} + \mathbb{R}_{44} 
\end{array}
\right) \Sigma^+\,.
\end{equation}

We can determine the corresponding non-geometric fluxes which will turn on these charge entries by analyzing the structure of the linear equations and keeping in mind the liberty to add terms perpendicular to all basis elements in the symplectic basis. We find the following $Q$-fluxes (are one-forms and bi-vectors)
\begin{equation}
\begin{aligned}
Q = 
& + {1\over 2}\mathbb{Q}_{21} \left( \Re (e^2)\wedge \Im (\partial_1) 
                - \Re (e^1) \wedge \Im (\partial_2) \right) \wedge \Re (\partial_3) \cr
&+ \mathbb{Q}_{22} \left ( - \Re (e^2) \wedge \Re (\partial_1) + \Im (e^2) \wedge  \Im (\partial_1)
                           - \Im (e^1) \wedge \Im  (\partial_2) + \Re (e^1) \wedge \Re (\partial_2)
       \right)  \wedge \Re (\partial_3) \cr
&+ {1\over 3} \mathbb{Q}_{24}  \left( 
        \Im (e^1) \wedge \Re (\partial_2) - \Im (e^2) \wedge \Re (\partial_1) 
       \right) \wedge \Re (\partial_3) \cr
& + \mathbb{Q}_{32} \Im (\partial_1) \wedge \Im (\partial_2) \wedge \Re (e^3)   \cr
& - {1\over 2} \mathbb{Q}_{33}   \Re (\partial_1) \wedge \Re (\partial_2) \wedge \Re (e^3)  \cr
& - {1\over 12} \mathbb{Q}_{34}   (  \Im (\partial_2) \wedge \Re (\partial_1) 
                     +  \Re (\partial_2) \wedge \Im (\partial_1))\wedge \Re (e^3)  \cr
& + 3 \mathbb{Q}_{42} \Im (\partial_1) \wedge \Im (\partial_2) \wedge \Im (e^3)   \cr
& - {3\over 2}  \mathbb{Q}_{43}       \Re (\partial_1) \wedge \Re (\partial_2) \wedge \Im (e^3)  \cr
& -{1\over 4} \mathbb{Q}_{44}    ( \Im (\partial_2) \wedge \Re (\partial_1) + \Re (\partial_2) \wedge  \Im (\partial_1)) \wedge \Im (e^3) \,,
\end{aligned}
\end{equation}
as well as $R$-fluxes of the type
\begin{equation}
\begin{aligned}
R 
=& - {1\over 8} \mathbb{R}_{31}  \Im (\partial_1) \wedge \Im(\partial_2) \wedge \Im (\partial_3) \cr
  & -{1\over 2}\mathbb{R}_{32}   (\Re (\partial_1) \wedge \Im (\partial_2) - \Re (\partial_2) \wedge \Im (\partial_1)) \wedge \Im (\partial_3) \cr
 & -{1\over 12} \mathbb{R}_{34}  \Re (\partial_1) \wedge \Re (\partial_2) \wedge \Im (\partial_3) \cr
  & +{3\over 8} \mathbb{R}_{41}  \Im (\partial_1) \wedge \Im (\partial_2) \wedge \Re (\partial_3) \cr
  & -{3\over 2} \mathbb{R}_{42}  (\Re (\partial_1) \wedge \Im (\partial_2) - \Re (\partial_2) \wedge \Im (\partial_1) ) \wedge \Re (\partial_3) \cr
  & + {1\over 4} \mathbb{R}_{44}  \Re (\partial_1) \wedge \Re (\partial_2) \wedge \Re (\partial_3)  \,.
\end{aligned}
\end{equation}

In summary we have shown that the full charge matrix can be constructed by switching on geometric as well as non-geometric fluxes:
\begin{equation}
\mathcal{Q}= 
\left(
\begin{array}{c|c|c|c}
\mathbb{F}_{11}  & \mathbb{F}_{12} +\mathbb{H}_{12} &  \mathbb{H}_{13}& \mathbb{F}_{14}+\mathbb{H}_{14} \cr \hline
 \mathbb{F}_{21} + \mathbb{Q}_{21} & \mathbb{F}_{22} + \mathbb{H}_{22}+ \mathbb{Q}_{22}& \mathbb{F}_{23} + \mathbb{H}_{23}&\mathbb{F}_{24}+ \mathbb{H}_{24} + \mathbb{Q}_{24}\cr \hline
\mathbb{R}_{31} & \mathbb{Q}_{32} + \mathbb{R}_{32} &\mathbb{Q}_{33}& \mathbb{Q}_{34} + \mathbb{R}_{34} \cr \hline
\mathbb{R}_{41}& \mathbb{F}_{42} + \mathbb{Q}_{42}+ \mathbb{R}_{42}& \mathbb{F}_{43} +\mathbb{Q}_{43} &  \mathbb{F}_{44}+  \mathbb{Q}_{44} + \mathbb{R}_{44} 
\end{array}
\right) \,.
\end{equation}


\subsubsection{Mirror symmetry}

We now wish to test out generalized SYZ proposal in this setup. This should in particular be compatible with the proposed mirror map of \cite{Grana:2006hr}.
The mirror fluxes are obtained by first recalling that we dualize along $\Re (e^1)$, $\Re (e^2)$ and $\Re (e^3)$ and that thereby the mirror map is realized as
\begin{equation}
\begin{aligned}
\Re (e^i) \quad  \longleftrightarrow  \quad \Re (\partial_i) \,.
\end{aligned}
\end{equation}
The mirror fluxes are determined straight-forwardly from our expressions for the fluxes. The mirrors of the  
geometric fluxes are
\begin{equation}
\begin{aligned}
\widehat{f}
    =& +2 \mathbb{F}_{11} \Re (e^3) \wedge \Re (\partial_1) \wedge \Re (\partial_2)\cr
      &+ 8 \mathbb{F}_{12} \left(\Re (e^3) \wedge \Im (e^1) \wedge \Re (\partial_2)  
              + \Re (e^3) \wedge \Re (\partial_1) \wedge \Im (e^2) \right) 
   \cr
   &+ {4\over 3} \mathbb{F}_{14} \Re (e^3) \wedge \Im (e^1) \wedge \Im (e^2)  
   \cr
      & - 2 \mathbb{F}_{21} \Im(\partial_3) \wedge \Re (\partial_1) \wedge \Re (\partial_2) \cr
     &-4 \mathbb{F}_{22} ( \Im (\partial_3) \wedge \Im(e^1) \wedge \Re (\partial_2)
                          +\Im (\partial_3) \wedge \Re(\partial_1) \wedge \Im (e^2) ) \cr 
     & - 2 \mathbb{F}_{22} ( \Im(\partial_1) \wedge \Re (\partial_2) \wedge \Im(e^3)
                            + \Im(\partial_2) \wedge \Im(e^3) \wedge \Re (\partial_1)) \cr  
     & + 2 \mathbb{F}_{23} (\Re (e^1)\wedge \Im (e^3) \wedge \Im (e^2)
                 + \Re (e^2) \wedge\Im(e^1) \wedge \Im (e^3) )\cr
     & - {1\over 3} \mathbb{F}_{24} \left(2 \Im (\partial_3)\wedge \Im(e^1) \wedge \Im(e^2) 
                + \Re (e^1)\wedge \Re (\partial_2) \wedge \Im(e^3)  
                  + \Re (e^2) \wedge \Im(e^3) \wedge \Re (\partial_1) \right)\cr
     & +  3 \mathbb{F}_{42} (- \Re (\partial_2) \wedge \Im (\partial_1) + \Re (\partial_1) \wedge \Im (\partial_2)) \wedge \Re (\partial_3) \cr
     &+ {3\over 2} \mathbb{F}_{43} (\Re (e^1) \wedge \Im (e^2) \wedge \Re (\partial_3) 
           + \Re (e^2) \wedge \Re (\partial_3) \wedge \Im (e^1)) \cr 
        & +{1\over 2} \mathbb{F}_{44} (\Im (e^2) \wedge \Im (\partial_1) + \Im (\partial_2) \wedge \Im (e^1)) \wedge \Re (\partial_3)  \,,
   \end{aligned}
\end{equation}
and
\begin{equation}
\begin{aligned}
\widehat{H} 
    =&       +  16 \mathbb{H}_{12} \Re (\partial_1) \wedge \Re (\partial_2) \wedge \Im (e^3) \cr
      &             -8 \mathbb{H}_{13} \Im (e^1) \wedge \Im (e^2) \wedge \Im (e^3)  \cr
   &  +  {4\over 3} \mathbb{H}_{14} \left(\Re (\partial_2) \wedge \Im (e^1 )- \Re (\partial_1) \wedge \Im (e^2) \right) \wedge \Im (e^3)  \cr
   &  +  16 \mathbb{H}_{22} \Re (\partial_1) \wedge \Re (\partial_2) \wedge \Re (\partial_3) \cr
    &    - 8 \mathbb{H}_{23} \Im (e^1) \wedge \Im (e^2) \wedge \Re (\partial_3) 
    \cr
   &  -  {4\over 3} \mathbb{H}_{24} \left(\Im (e^1 )\wedge \Re (\partial_2) \wedge \Re (\partial_3) 
      + \Re (\partial_1) \wedge \Im (e^2) \wedge \Re (\partial_3) \right) \,.
\end{aligned}
\end{equation}
These include of course both geometric and non-geometric fluxes. 

Likewise the non-geometric mirrors are
\begin{equation}
\begin{aligned}
\widehat{Q}
 =
& + {1\over 2}\mathbb{Q}_{21} \left( \Re (\partial_2)\wedge \Im (\partial_1) 
                - \Re (\partial_1) \wedge \Im (\partial_2) \right) \wedge \Re (e^3) \cr
&+ \mathbb{Q}_{22} \left ( - \Re (\partial_2) \wedge \Re (e^1) + \Im (e^2) \wedge  \Im (\partial_1)
                           - \Im (e^1) \wedge \Im  (\partial_2) + \Re (\partial_1) \wedge \Re (e^2)
       \right)  \wedge \Re (e^3) \cr
&+ {1\over 3} \mathbb{Q}_{24}  \left( 
        \Im (e^1) \wedge \Re (e^2) - \Im (e^2) \wedge \Re (e^1) 
       \right) \wedge \Re (e^3) \cr
& + \mathbb{Q}_{32} \Im (\partial_1) \wedge \Im (\partial_2) \wedge \Re (\partial_3)   \cr
& - {1\over 2} \mathbb{Q}_{33}   \Re (e^1) \wedge \Re (e^2) \wedge \Re (\partial_3)  \cr
& - {1\over 12} \mathbb{Q}_{34}   (  \Im (\partial_2) \wedge \Re (e^1) 
                     +  \Re (e^2) \wedge \Im (\partial_1))\wedge \Re (\partial_3)  \cr
& + 3 \mathbb{Q}_{42} \Im (\partial_1) \wedge \Im (\partial_2) \wedge \Im (e^3)   \cr
& - {3\over 2}  \mathbb{Q}_{43}       \Re (e^1) \wedge \Re (e^2) \wedge \Im (e^3)  \cr
& -{1\over 4} \mathbb{Q}_{44}    (\Im (\partial_2) \wedge \Re (e^1) + \Re (e^2) \wedge  \Im (\partial_1)) \wedge \Im (e^3) \,,
\end{aligned}
\end{equation}
and
\begin{equation}
\begin{aligned}
\widehat{R} 
=& - {1\over 8} \mathbb{R}_{31}  \Im (\partial_1) \wedge \Im(\partial_2) \wedge \Im (\partial_3) \cr
  & - {1\over 2}\mathbb{R}_{32}   (\Re (e^1) \wedge \Im (\partial_2) - \Re (e^2) \wedge \Im (\partial_1)) \wedge \Im (\partial_3) \cr
 & -{1\over 12} \mathbb{R}_{34}  \Re (e^1) \wedge \Re (e^2) \wedge \Im (\partial_3) \cr
  & +{3\over 8} \mathbb{R}_{41}  \Im (\partial_1) \wedge \Im (\partial_2) \wedge \Re (e^3) \cr
  & -{3\over 2} \mathbb{R}_{42}  (\Re (e^1) \wedge \Im (\partial_2) - \Re (e^2) \wedge \Im (\partial_1) ) \wedge \Re (e^3) \cr
  & + {1\over 4} \mathbb{R}_{44}  \Re (e^1) \wedge \Re (e^2) \wedge \Re (e^3)  \,.
\end{aligned}
\end{equation}

Acting with the mirror fluxes on the basis yields the mirror charge matrix $\widehat{\mathcal{Q}}$ to be
\begin{equation}
\widehat{\mathcal{Q}}
=
\left(
\begin{array}{c|c|c|c}
-{1\over 4} \mathbb{Q}_{33} & {3\over 16} \mathbb{F}_{43} - {3\over 32} \mathbb{Q}_{43}& \mathbb{H}_{13}& -6 \mathbb{F}_{23} - 6 \mathbb{H}_{23}  \cr\hline
 -{1\over 6} \mathbb{Q}_{34} + {1\over 24} \mathbb{R}_{34}& {1\over 8 }\mathbb{F}_{44} + {1\over 8} \mathbb{Q}_{44} + {1\over 16} \mathbb{R}_{44}& -{1\over 6 } \mathbb{F}_{14} + {2\over 3  }  \mathbb{H}_{14}&  - \mathbb{F}_{24} - 2 \mathbb{H}_{24} +{1\over 4} \mathbb{Q}_{24} \cr\hline
 \mathbb{R}_{31}& {3\over 8} \mathbb{R}_{41} & - 4 \mathbb{F}_{11} & -24 \mathbb{F}_{21} - 6 \mathbb{Q}_{21} \cr\hline
 {8\over 3} \mathbb{Q}_{32} + {2\over 3} \mathbb{R}_{32}& -2 \mathbb{F}_{42} + \mathbb{Q}_{42} + {1\over 2} \mathbb{R}_{42}& {16\over 3} \mathbb{F}_{12} - {32 \over 3} \mathbb{H}_{12} & 16 \mathbb{F}_{22} - 64 \mathbb{H}_{22} + 2 \mathbb{Q}_{22} 
\end{array}
\right) \,.
\end{equation}
Note this is nicely confirming the conjectured mirror map on the charge matrix as of \cite{Grana:2006hr} where it was conjectured that the charge entries appear as 
\be
\mathcal{Q} = \left(  \begin{array}{cc}    
p_I{}^A & e_{IB} \cr
q^{IA} & m^I{}_B
\end{array}
\right)     
\quad \rightarrow \quad 
\widehat{\mathcal{Q}}
 = \left(  \begin{array}{cc}     
m^I{}_A & e_{BI} \cr
q^{AI} & - p_I{}^B
\end{array}
\right) \,.    
\ee
Recall that this was derived by comparing the Killing prepotentials, and thus does not fix the mapping of the charges up to linear transformations that leave the blocks invariant. We confirmed the mapping of the charges and explicitly worked out the charge entries of $\widehat{\mathcal{Q}}$.

 We should note that in addition to the linear conditions that arise from the action of the fluxes on the basis, there are also quadratic constraints, which arise from the condition that the differential has to be nilpotent, upon the entries of the charge matrix. These will have to be taken into account, in order to discuss physical flux configurations. The factors in the above matrix could then be taken care of by allowing only fluxes that solve the quadratic constraints.


\section[Supersymmetric cycles on generic $SU(3)\times SU(3)$  structure backgrounds]{Supersymmetric cycles on generic $SU(3)\times SU(3)$ \\ structure backgrounds}

In this section we want to investigate the generic case of  $SU(3)\times SU(3)$ structure backgrounds, where the underlying  manifold (in some duality frame) has non-zero Euler number. Relaxing the topological condition $\chi(M)=0$ implies that there is no static $SU(2)$ structure at all. Not only do we have to face the loss of ordinary complex structure on $M$, but we are going to encounter type-jumping phenomena. The following two closed subsets are indeed going to be of special interest:
\be
\{ c^{-1}(1) \} \; : \qquad \hbox{type-three and type-zero pure spinors}\,,
\ee
as in the case of $SU(3)$ structures (this set was empty in the
previous part of our analysis), and
\be
\{ c^{-1}(0) \} \;: \qquad \hbox{type-one and type-two pure spinors} \,,
\ee
as in the case of $SU(2)$ structures. They correspond to the two big
circles we have depicted on figure (\ref{fig1}). So far we have been
confined to only one of them, because of the topological assumption we
have made.

 Some three-tori will be supersymmetric on $M\times \hat{M}$, either
 in transverse or longitudinal position, but they will always be
 situated above points of these two special subsets. Away from those
 subsets, types of pure spinors are too low to allow for stable
 D0-branes. This is T-dual to the disappearance of most of the
 supersymmetric three-torus fibers. We shall describe this in
 terms of mass generation for moduli through fluxes.

 Motivated by this observation concerning D0-branes, we want to
 address the existence, stability and moduli space of
 three-dimensional supersymmetric cycles. We shall see that for
 $SU(3)\times SU(3)$ structures that are not static $SU(2)$
 structures, such cycles still exist at type-jumping points.  As
 T-duality does not change moduli spaces, we expect some moduli of
 those cycles to be fixed. In particular, three-dimensional
 supersymmetric cycles are not likely to give rise to a fibered
 structure of a whole manifold with generic $SU(3)\times SU(3)$
 structure. But they can still allow for the exchange of pure spinors
 $\Phi_-$ and $\Phi_+$ by mirror symmetry as a T-duality along a
 three-dimensional supersymmetric cycle.


\subsection{D3-branes and D0-branes through maximum-type points}

Consider a mirror pair of manifolds $M$ and $\hat{M}$ with
$SU(3)\times SU(3)$ structures, that do not fall into the class of
static $SU(2)$ structures (as they have opposite Euler
characteristics, assuming that one has non-zero Euler characteristic
is sufficient to ensure the condition). We assume both sides of
the mirror correspondence to have a geometric description in the sense
of a sigma model. 
Consider some point $p$ on $M$ at which the complex
one-form $v+iw$ vanishes. At that point the pure spinors assume the
same forms as in the Calabi--Yau case. We may write for some complex coordinates $X,Y,Z$  
\be
\ba
{\Phi_-}|_p=\Omega|_p &=dX\wedge dY\wedge dZ \cr
\label{typezero}
{\Phi_+}|_p=e^{iJ}|_p &=e^{\frac{i}{2}(dX\wedge d\bar{X}+dY\wedge d\bar{Y}+dZ\wedge d\bar{Z})}  \,,
\ea
\ee
and one may put a D0-brane of
the B-model, that is generalized complex
w.r.t. to $\Omega$, or a D3-brane of the A-model, {\it i.e.} a
Lagrangian D-brane which will be denoted by $L$.


 Let us T-dualize along $L$, which we assume to have the topology of a
 torus, corresponding to the three isometries we need to perform
 T-duality\footnote{The assumption is reasonable because we have two $SU(3)$ structures, each of which gives rise to a fibration by three-tori, and at points the two fibers are the same, the fiber is supersymmetric; but such points are exactly the zeroes of $v+iw$.}. Let there be local coordinates $x'$,$y'$ and $z'$ on $L$
 (that are imaginary parts of complex coordinates $X=x+ix'$,
 $Y=y+iy'$, $Z=z+i z'$ defined on the locus with equation $v+iw = 0$),
 so that Fourier--Mukai transform yields
\be
 \ba 
 \ft\left(\Phi_-|_{\ft(L)}\right) & =  \int_L e^{i(dx'\wedge d\hat{x'}+
 dy'\wedge d\hat{y'} +dz'\wedge d\hat{z'})} dX\wedge dY\wedge dZ \cr
& =  \exp\left( i(d\hat{x'}\wedge dx + d\hat{y'}\wedge
 dy + d\hat{z'}\wedge dz)\right)\cr
&  = \hat{\Phi}_+|_{\hat{p}} =: e^{i\hat{J}}|_{\hat{p}} \cr
\ft\left(\Phi_+|_{\ft(L)}\right)& = \hat{\Phi}_-|_{\hat{p}} =: \hat{\Omega}|_{\hat{p}} \,,
\ea
\ee
which are the expressions of the pure spinors on the T-dual point $\hat{p}$ on which a supersymmetric D0-brane sits. Of course $\hat{p}$ has to be in the set of zeroes of $\hat{v}+i\hat{w}$, or $\hat{c}^{-1}(1)$, which is not empty since the mirror manifold $\hat{M}$ also has non-zero Euler number.

\subsection{Away from  maximum-type points through fluxes}

It has long been appreciated that the behaviour of pure spinors under mirror symmetry is transparent to $B$-transforms by a two-form whose components are extended in directions transverse to the T-dualized directions, while non-geometry occurs when the two-form has components that are longitudinal. In terms of the previous local complex coordinates, $B$-transforms by two-forms of type $(1,1)$ are still $B$-transforms on the mirror, while those of type $(0,2)$ or $(2,0)$ are $\beta$-transforms on the mirror. This can be seen in local charts by wedging together pairs of the following naturality properties derived in lemma 6.2 of the second reference in \cite{BB1}, where $v$ and $\phi$ denote a longitudinal vector and one-form, and $w$ and $\psi$ denote a transverse vector and one-form:
\begin{eqnarray}
i)&\;\ft(\iota_v \wedge \Phi)&= \hat{v}\wedge (\ft (\Phi)) \nonumber \\
ii)&\;\ft(\iota_w \wedge \Phi)&= \iota_w (\ft (\Phi)) \nonumber \\
iii)&\;\ft(\phi\wedge \Phi) &= \iota_{\hat{\phi}} (\ft(\Phi)) \nonumber \\
iv)&\;\ft(\psi\wedge \Phi)   &= \psi\wedge \ft(\Phi) \,.\nonumber 
\end{eqnarray}
In other words covariant and contravariant  tensors stay so under T-duality if their components are transverse to the dualized directions, while they are flipped if they are longitudinal.

So far we have seen how T-duality maps pure spinors $\Phi_+$ and $\Phi_-$ to
each other along the maximum-type locus of equation $v+iw=0$. It looked formally the same as in the Calabi--Yau or $SU(3)$ structure case. Suppose
an $H$-flux is turned on on both sides of the mirror correspondence. Choosing a gauge for the local $B$-field from which the flux derives induces various $B$- and
$\beta$-transforms on $M$ and $\hat{M}$, according to the way the support of the $B$-field intersects with the T-dualized directions. Generically, going away from the maximum-type locus should induce a $\beta$-transform that will lower the type of $\Phi_-$ to one, which is the most generic type for an odd pure spinor (\ie\ the lowest type allowed by parity).

Thanks to property $iv)$, $B$-fields of type $(1,1)$ in the complex structure described above pull back to zero on the three-cycle $L$. They act as $B$-transforms on  both sides of the  mirror correspondence and do not lower the type of the pure spinors
\be 
e^B\wedge\Phi_\pm \longleftrightarrow e^B\wedge{\hat{\Phi}}_\mp \,.
\ee
 
We have to take into account possible $B$-transforms by longitudinal $B$-fields, that give rise to non-geometric fluxes on the mirror (we will restrict to the case of $Q$-fluxes on the mirror, with two indices of the $B$-field along T-dualized directions). Consider an $H$-flux on the space $M$, with one unit of flux along a three-cycle $\mathcal{C}$:
\be 
\int_\mathcal{C} H=1\,.
\ee
In order for $L$ to be a supersymmetric three-cycle, the local
$B$-field, which gives rise to the flux, has to pull-back to zero on $L$.

We are interested in the application of T-duality in two directions
carrying indices of non-zero components of the $H$-flux. These directions,
denoted by $y$ and $z$, are
two $U(1)$ isometries, spanning a two-torus. Consider the one-form
valued integral of $H$ along this torus. It is closed because the
three-form $H$ is: \be d\int_{T^2} H=0.\ee  
One can locally integrate the one-form, so that there exists (locally) a
scalar function $X$ such that 
\be 
\int_{T^2} H= dX \,,
\ee
 which amounts to a
gauge choice, because the $B$-field 
\be 
B:= X {\mathrm{vol}}_{T^2} \,,
\ee 
where $\mathrm{vol}_{T^2}$ is the volume form of $T^2$,
is compatible
with the quantization of $H$.
Upon T-duality along the two $U(1)$ isometry directions $y$ and $z$, this
$B$-field is mapped to a bivector living on the $T$-dual manifold
\be 
\ft(
e^{i(J+Xdy\wedge dz)})=
X\iota_{\partial_{\hat{y}}} \iota_{\partial_{\hat{z}}}
        \left(  \ft(e^{iJ})\right) + \ft(e^{iJ})\,.
\ee 

This way the lowest component of the odd pure spinor we read-off from the 
{RHS} 
is the one-form
  $X\iota_{\partial_{\hat{y}}}\iota_{\partial_{\hat{z}}}\Omega$, that appears to be weighted
  by the local coordinate $X$. We may thus identify the first term in 
  the expansion of the polyform on the {\sc{RHS}} with the mirror of
  the fiberwise components of the $H$-field. It should also be equal to the
  one-form $v+iw$. 
Thus, we have seen that the $(0,2)$ part of the argument of the
exponential in the expression of $\Phi_+$ is mirror to the complex
vector $v+iw$.  We can rewrite this mapping in a coordinate-independent way as
\be
\Phi_+=e^{B^{(0,2)}}\wedge e^{iJ}\longrightarrow e^\beta\hat{\Omega}=\Phi_-,
\ee
where the {\sc{RHS}} has now type one and contains an overall factor of $\hat{v}+i\hat{w}$.

Likewise we can start with $\Phi_-$. Again $H$ can be locally written as $H= d(X dy \wedge dz)$ and thus
\be 
\ft (e^{X dy \wedge dz} \wedge \Omega) 
=  e^{X \partial_{y} \wedge \partial_{z}} \ft (\Omega) \,.
\ee
Furthermore $\ft (\Omega) = e^{i J}$ with $J= J^{1,2} = j \pm v\wedge w$ being one of the two-forms of the two $SU(3)$ structures (\ref{JJDef}). We have thus established that
\be 
e^{B^{(0,2)}}\wedge\Omega \longrightarrow e^\beta e^{iJ} \,,
\ee 
with $\beta = X \partial_{y} \wedge \partial_{z}$.  
 Apart from this we know that the contraction between $\beta$ and $j$ vanishes, just as $\omega\wedge j$ vanishes for $SU(2)$ structures. Hence contractions between the bivector
$\beta$ and the polyform $e^{iJ}$ only involve $j$ and is therefore
unambiguous. 
So the contraction between $\beta$ and the higher powers of $J$
just selects the square of $j$, and gives rise to a $(2,0)$ form
called $B'$, which squares to zero. Using the expansion (\ref{typezero}) we find:
\be
e^\beta(e^{iJ})=1+iJ+\beta\left(-\frac{J^2}{2}\right)-\frac{J^2}{2}+\beta\left(-i\frac{J^3}{6}\right)-i\frac{J^3}{6} \,.
\ee
Defining
\be
B':=\beta\left(-\frac{J^2}{2}\right) \,,
\ee
this can be rewritten in the following way
\be
e^\beta\left(e^{iJ}\right)=1+iJ+B'+B'\wedge iJ-\frac{J^2}{2}-i\frac{J^3}{6}=e^{B'}\wedge e^{iJ}\,.
\ee
Thus, one may say that the $\beta$-transform of the type-zero pure spinors assumes the same form as a $B$-transform for accidental dimensional reasons. We therefore write the $\beta$-transform of the type-zero
spinor as a $B$-transform by $B'$, which of course is still of the most
generic type zero:
\be 
e^{B^{(0,2)}}\wedge\Omega  \longrightarrow e^{B'+iv\wedge w + ij} \,.
\ee
This  formula was already derived assuming a $T^3$-fibration in
\cite{Grana:2006kf} as a clue that $SU(3)\times SU(3)$ structures
could account for non-geometric situations involving T-dualizing with
a $B$-field extended along two fiberwise directions. Here we see that
it actually holds for a mere topological reason on spaces with
non-zero Euler number and $SU(3)\times SU(3)$ structure. On such
spaces $v+iw$ have zeroes on which odd pure spinors have type three,
thus giving rise to supersymmetric three-cycles; the mirror formula
between $\Phi_-$ and $\Phi_+$ follows from the naturality properties
of $B$- and $\beta$-transforms w.r.t. T-duality along the
three-cycles, even if the SYZ argument is spoiled away from the zeroes
of $v+iw$ due to the absence of supersymmetric D0-branes. Moreover,
T-dualizing along $L$ and exploiting properties of the Fourier--Mukai
transform allowed us to go the other way around, which lowers the type
of $\Phi_-$. To sum up, putting all the possible $B$- and
$\beta$-transforms on both sides, we have argued that the following
T-duality holds in an open neighborhood of type-jumping point:
\begin{eqnarray}
 \Phi_+=e^{\tilde{\beta}+{B'}^{(0,2)}+B^{(1,1)}}\exp(iJ) \qquad  &\longleftrightarrow &\qquad  e^{\beta'+\tilde{B}^{(0,2)}+B^{(1,1)}}\hat{\Omega} =\hat{\Phi}_- \nonumber\\
\tilde{\beta}  \qquad  &\longleftrightarrow &\qquad  {\tilde{B}^{(0,2)}}  \nonumber\\
{B'}^{(0,2)}  \qquad  &\longleftrightarrow &\qquad  \beta' \nonumber\\
B^{(1,1)}   \qquad  &\longleftrightarrow &\qquad B^{(1,1)}\,,
\end{eqnarray}
where the odd pure spinor has type one as $\beta'$ (or equivalently ${B'}^{(0,2)}$) is non-zero.


\subsection{Moduli spaces}

As D0-branes can only be stable at points where the odd pure
spinors has type three, their moduli space must be evaluated by
looking at massless infinitesimal deformations at such exceptional
points. Going away from such a point involves a $\beta$-transform. If one goes along the subset $c^{-1}(1)$ the $\beta$-transform is trivial and we have found a translation modulus; otherwise the direction along which we are going is a fixed modulus. On the other hand, the first cohomology of the Lie algebroid of a D0-brane was
evaluated in \cite{Koerber:2006hh} as the set of vectors $X^l$ such
that the $\beta$-transform that acts on the pure spinor satisfies
\be 
\partial_l\beta^{\mu\nu} X^l \partial_\mu\wedge\partial_\nu=0 \,.
\ee
This makes for a five-dimensional moduli space, as a gauge may be
chosen in which $\beta$ only depends on one coordinate, the one along
the direction $v$. Consider the three-dimenional D-brane going through
such a point. It also has a five-dimensional moduli space, since the normal deformation in the direction $v$ is not allowed anymore, and it is exactly the modulus that has disappeared for D0-branes.

 In the more generic cases we want to investigate here, we have to compute
 the mass matrix of the deformations of our three-dimensional supersymmetric
 tori. Moduli that are fixed by the flux should get a mass.

 One can make an observation in local coordinates around a point where
 the pure spinors have type zero and type three. The fundamental
 two-form $J$ takes the expression 
\be 
J=\frac{i}{2}\left(dX\wedge d\bar{X} +dY\wedge d\bar{Y} + dZ\wedge d\bar{Z}\right)\,,
\ee 
and
 imagine we start with a supersymmetric three-torus extended along the
 directions $x$, $y$ and $Z$ and the T-dual of an $H$-flux deriving
 from the coordinate $x$ is a $\beta$-transform with
 $\beta=X\partial_y\wedge\partial_z$, so that it is easy to repeat the argument of the previous subsection for the computation of $B'$. In a neighborhood of the point we considered, $\Phi_+$ assumes the
 form: 
\be 
e^\beta e^{iJ}=e^{X d\bar{Y}d\bar{Z}} e^{iJ}\,,
\ee 
so that a
 four-chain $\mathcal{B}$ that is bounded by the supersymmetric
 three-cycle and some generalized cycle $(\Sigma,F)$ at the other end
will go (along the $X$
 direction)
 through cycles carrying non-zero field strength
\be F=
 P_{\Sigma}(X d\bar{Y}d\bar{Z}),
\ee 
where $P_{\Sigma}$ denotes the pull-back to $\Sigma$. Hence the three-cycle $(\Sigma,F)$ cannot be generalized complex if it
 goes into the $X$ direction. This loss of structure fixes the position moduli $X$ for the three-cycle, which fact is mirror to $X$ acquiring a mass as a translation modulus for a D0-brane.


\section{Conclusions and outlook}

The SYZ argument has been shown to extend to a class of generalized Calabi-Yau spaces, namely so-called static $SU(2)$ structure manifolds. We have shown that there are no supersymmetric three-tori on $M$ or its mirror
 $\hat{M}$, but the product $M\times\hat{M}$ is doubly fibered by
 three-tori, both families of fibers are transverse to $M$ and
 $\hat{M}$, and the resulting six-tori are calibrated generalized
 submanifolds of $M\times \hat{M}$. Moreover mirror symmetry is
 performed by T-dualizing the three-dimensional intersection of such
 generalized submanifold with $M$. This transversality property is
 reminiscent of the (much more general) conjectures formulated by
 Gualtieri in the final chapter of \cite{Gualtieri:2003dx}.

It is somewhat surprising that this argument is applicable also when including non-geometric fluxes, 
in particular $R$-fluxes. These non-geometric fluxes are expected to spoil the geometric description of the background even locally. In the $R$-flux case, the geometry is expected to be replaced by some non-associative 
algebra \cite{Bouwknegt:2004ap}. However we did not 
encounter such a necessity. We suspect that the case of static $SU(2)$ structure, which prevents 
the type of the pure spinors from jumping, guards us against the destabilizing effects of non-geometric 
fluxes on D-branes. 

The large-volume limit which was assumed in the SYZ argument for Calabi--Yau manifolds is also highly questionable in generic flux backgrounds. Again the topological condition of a static $SU(2)$ with a non-vanishing vector field allows for more globally well-defined quantities than the ordinary complex torus studied in \cite{Lawrence:2006ma,Grange:2006es}. This is consistent with the observation made in \cite{Grana:2006hr}
 that more charges can be turned on geometrically on $SU(2)$ structure backgrounds than on generic ones.

The case of
generic $SU(3)\times SU(3)$ structures is much less transparent.\footnote{We have disregarded Ramond--Ramond fluxes, in the presence
of which a one of the two pure spinors cannot be closed
\cite{Grana:2005sn}. Bianchi identity in the presence of Ramond--Ramond fluxes  requires an orientifold projection, see \cite{Benmachiche:2006df}.}
 We have identified a set of three-cycles, T-dual to type-jumping points on the mirror. They cannot fiber the manifold or even its product with its mirror. This fact is mirror to the mass that fluxes give to the translation moduli of D0-branes, spoiling the very first step of the SYZ argument. T-dualizing the surviving three-tori and asking for functorial properties w.r.t. $B$- and $\beta$-transforms of generalized geometry gives however a correct mirror exchange between type-zero and type-one pure spinors. Our argument was limited to the use of classical geometry.

In order to formulate an SYZ argument for the generic case,  it seems natural to consider non-commutative fibrations.
 It has been observed that T-dualizing directions that support more than one index of a non-zero component of a $B$-field leads to non-commutative fibrations through an {\emph{uncertainty principle}} for D-branes \cite{Kapustin:2003sg,Grange:2004ra}. Of course allowing non-commutative fibers, with non-commutativity scale proportional to the quanta of fluxes and to the discrepancy between the pair of $SU(3)\times SU(3)$ structures, would be a way of fibering generalized backgrounds by (further) generalized submanifolds. The only fibers we are able to see in the present approach are the ones along which the two structures agree, which results in type-jumping and in a commutative fiber. It might be that non-commutative fibrations on more general bases than a torus will be equivalent to fibrations by T-folds \cite{Hull:2004in}, and that going away from type jumping points will require acting on the fibers with transition functions involving T-dualities.

We hope to gain more insight into these issues by studying the proper
reduction on generic $SU(3) \times SU(3)$ structure manifolds. Initial
results have appeared in \cite{Grana:2006hr} and the case of $SU(3)$
structures was discussed in \cite{KashaniPoor:2006si}. We trust that
the analog of harmonic forms will be generalized or twisted harmonic
forms, {\it i.e.} forms that are harmonic w.r.t. the Laplacian twisted
by all fluxes (geometric and non-geometric). This should in particular
allow one to determine the mass terms that we discussed in at the end
of this paper, and thus the disappearance of geometric moduli will
become more transparent. We shall come back to these points in due
time.


\subsection*{Acknowledgments}

We thank Mariana Gra$\tilde{{\rm n}}$a, Jan Louis, Luca Martucci, Ron Reid-Edwards, Bastiaan Spanjaard and Jie Yang for discussions and useful correspondence. P.G. is funded by the German-Israeli Foundation for Scientific Research and Development. S.S.N. is funded by a Caltech John A. McCone Postdoctoral Fellowship in Theoretical Physics and thanks Jan Louis for hospitality at the II. Institut f\"ur theoretische Physik of the University of Hamburg. 
This work was supported in part by the DFG and the European RTN Program MRTN-CT-2004-503369.


\bibliographystyle{JHEP} \renewcommand{\refname}{References}

\providecommand{\href}[2]{#2}\begingroup\raggedright\endgroup


\end{document}